\documentclass[amsmath,eqsecnum,preprint,pre,aps,showpacs,floatfix]{revtex4}
\usepackage{graphicx} 
\usepackage{amsmath}
\usepackage{graphicx}
\usepackage{float}
\usepackage{subfigure}
\usepackage{xcolor}
\usepackage{bm}

\begin{document}


\title{Non-Equilibrium Statistical Mechanics of Glass-Forming Liquids under Continuous Cooling: Microscopic Inner Clocks, Fictive Temperatures, and Kauzmann Diagrams}



\author{R. Peredo-Ortiz}
\affiliation{Instituto de F\'{\i}sica, Universidad Aut\'{o}noma de San Luis Potos\'{\i}, \'{A}lvaro Obreg\'{o}n 64, 78000 San Luis Potos\'{\i}, SLP, M\'{e}xico}

\author{M. Medina-Noyola}
\affiliation{Instituto de F\'{\i}sica, Universidad Aut\'{o}noma de San Luis Potos\'{\i}, \'{A}lvaro Obreg\'{o}n 64, 78000 San Luis Potos\'{\i}, SLP, M\'{e}xico}

\author{L. F. Elizondo-Aguilera}
\affiliation{Independent Researcher}

\date{\today}

\begin{abstract}
        The description of macroscopic properties in out-of-equilibrium amorphous solids such as gels and glasses, typically processed under finite cooling rates, remains a fundamental challenge for modern statistical physics. 
        In this work, we employ the Non-Equilibrium Self-Consistent Generalized Langevin Equation (NESCGLE) theory to construct a microscopic description of the structural and calorimetric  glass transition under continuous cooling. 
        Building upon the microscopic Onsager--Machlup interpretation of Narayanaswamy's material time (the inner clock $u(t)$) [J.~Chem.~Phys.\ \textbf{156}, 244506 (2022)], we recover Tool's fictive temperature $T_f$ directly from the evolving spatial pair correlations and show that wavenumber dispersion naturally generalizes the classical scalar $T_f$ into a continuous mode spectrum governed by de Gennes narrowing at the primary cage peak. 
        Across wide variations in cooling rate down to the quasi-static limit, the theory yields first-principles non-equilibrium Kauzmann diagrams for the two-body excess entropy that dynamically avert the Kauzmann crisis above the ideal dynamic arrest temperature $T_c$, while predicting sigmoidal relaxational heat-capacity steps consistent with computer simulations and calorimetry. 
        Finally, we extract the cooling-rate-dependent glass transition temperature $T_g(q)$ from both equal-area integrals and differential heat-capacity landmarks, demonstrating that its asymptotic approach toward $T_c$ obeys a critical power-law scaling governed directly by the equilibrium dynamic arrest exponent.             
    \end{abstract}



\maketitle

\section{Introduction}\label{sec:Introduction}

When a glass-forming liquid is cooled sufficiently fast below its equilibrium melting, it bypasses crystallization and enters a metastable supercooled regime~\cite{Kauzmann1948, Debenedetti2001, Angel1995, Berthier2009, GibaudSchurtenberguer}. 
Upon further cooling at a constant finite rate $q \equiv -\mathrm{d}T/\mathrm{d}t$, the intrinsic structural ($\alpha$-)relaxation time $\tau_\alpha(T)$ grows by many orders of magnitude until molecular or colloidal rearrangements can no longer keep pace with the external thermal ramp~\cite{Das2019, Ruta2013, Li2011, Lunkenheimer2023, hutchinson1995, bellon2000, Cangialosi2023}. 
At a cooling-rate-dependent kinetic glass transition temperature $T_g(q)$, the liquid falls out of thermodynamic equilibrium and vitrifies into an amorphous solid whose microstructure and macroscopic physical properties depend permanently on its thermal history~\cite{Tool1946, Kauzmann1948, DaviesJones1953, Kovacs1963, Narayanaswamy1971, Moynihan1976, KAHR1979, Scherer1984, Cooper1985, Bruning1992, Hodge1994, Vollmayr1996, Weyer2005, Gujrati2009, Yue2004, XinLi2017, Bravenec2023, Mauro2009, McKenna2020}. 
Macroscopically, this kinetic arrest is canonically synthesized in Kauzmann diagrams, where excess thermodynamic state functions---such as configurational entropy, enthalpy, and specific volume---are plotted against temperature~\cite{Kauzmann1948, GibbsDiMarzio1958, AdamGibbs1965, Debenedetti2001, Angel1995, Mauro2009, McKenna2020}.
Extrapolating the equilibrium supercooled branch toward low temperatures points to an entropy crisis at a finite Kauzmann temperature $T_K > 0$, where excess configurational entropy would vanish. 
Any finite cooling rate $q > 0$, however, kinetically averts this catastrophe by arresting the liquid at $T_g(q) > T_K$ onto a glassy plateau with frozen residual properties~\cite{Kauzmann1948, GibbsDiMarzio1958, AdamGibbs1965, Debenedetti2001, Mauro2009, McKenna2020, Lunkenheimer2023, Cangialosi2023}.

Because a glass violates the equilibrium equations of state, its macroscopic physical state cannot be uniquely specified by the instantaneous laboratory bath temperature $T(t)$ and density alone. 
Recognizing this fundamental limitation in 1946, Tool~\cite{Tool1946} introduced an internal structural state variable---the \emph{fictive temperature} $T_f(t)$---defined as the temperature at which the liquid in true thermodynamic equilibrium would exhibit the exact same structural state as the out-of-equilibrium material possesses at time $t$~\cite{Tool1946, DaviesJones1953}. 
A quarter-century later, Narayanaswamy~\cite{Narayanaswamy1971} resolved the simultaneous non-linearity and history dependence (``memory effect'') of structural recovery across the glass transformation range~\cite{Kovacs1963, Narayanaswamy1971, KAHR1979, Scherer1984, Hodge1994} by postulating that non-equilibrium relaxation becomes linear and time-translationally invariant when expressed not in terms of the laboratory clock $t$, but in terms of an internal ``material time'' (or inner clock) $\xi(t) = \int_0^t \mathrm{d}t'/\tau_\alpha[T(t'), T_f(t')]$~\cite{Narayanaswamy1971, Dyre2015, Dyre2018}. 
Shortly thereafter, Moynihan and coworkers~\cite{Moynihan1976} connected this formalism to calorimetry by identifying the normalized relaxational specific heat across the glass transformation range with the temperature derivative of the fictive temperature, $\tilde{C}_p(T) \equiv (C_p - C_{pg})/(C_{pe} - C_{pg}) = \mathrm{d}T_f/\mathrm{d}T$, and establishing the equal-area integral construction that relates the calorimetric glass transition temperature $T_g(q)$ directly to the limiting glassy fictive temperature $T'_f(q)$~\cite{Moynihan1976, Hodge1994, Weyer2005, Gujrati2009}.

Despite the broad empirical success of the resulting Tool--Narayanaswamy--Moynihan (TNM) phenomenology across silicate, metallic, polymeric, molecular, and colloidal glass formers~\cite{Tool1946, Narayanaswamy1971, Moynihan1976, KAHR1979, Scherer1984, Hodge1994, Weyer2005, Gujrati2009, Yue2004, XinLi2017, Bravenec2023, Mauro2009, McKenna2020, Svoboda2023}, its classical formulation remained disconnected from first-principles statistical mechanics: 
it relied on empirical Kohlrausch--Williams--Watts (KWW) stretched-exponential memory kernels and postulated a scalar fictive temperature $T_f(t)$, leaving unexplained why distinct macroscopic observables (such as enthalpy, volume, refractive index, or diffraction peak intensity) yield different fictive temperatures for the exact same thermal history~\cite{XinLi2017, McKenna2020, Cangialosi2023, DiLisio2023}.
Conversely, microscopic theories of glassy dynamics---such as Mode-Coupling Theory (MCT)~\cite{Goetze2009, Fuchs1999, Latz2000, Gory2019} and the Self-Consistent Generalized Langevin Equation (SCGLE) theory~\cite{Yeomans2000, Juarez2007, Pedro2011}---were inherently restricted to stationary equilibrium states. 
The extension to genuine out-of-equilibrium structural relaxation and physical aging was achieved by the Non-Equilibrium Self-Consistent Generalized Langevin Equation (NESCGLE) theory~\cite{Pedro2010, Pedro2010_2, LuisEnrique2013, Paty2017, Edilio2018, Edilio2019, Peredo2021, Gory_prl2022, Mendoza2022}, which is rooted in the non-equilibrium Onsager--Machlup theory of thermal fluctuations~\cite{Onsager1931_1, Onsager1931_2, Onsager1953, Machlup1953, Keizer1987, Peredo2024, MedinaNoyola2026AnnRev} and was initially applied to idealized instantaneous quenches ($q \to \infty$)~\cite{Pedro2010, Pedro2010_2, LuisEnrique2013, Paty2017, Mendoza2022}. 
Directly from the canonical relaxation equations of the non-equilibrium Onsager--Machlup theory, Peredo-Ortiz \textit{et al.}~\cite{Peredo2022} demonstrated from first principles that the TNM linear-convolution formalism and Narayanaswamy's internal material clock emerge automatically as the time integral of the state-dependent relaxation rate.
This theoretical connection between microscopic non-equilibrium statistical mechanics and the phenomenological material clock has been recognized across diverse contemporary contexts---including perspectives on slow dynamics in disordered materials~\cite{DelGado2023}, studies of multi-mechanism equilibration, physical aging, and size-dependent vitrification in polymeric, molecular, and metallic glasses~\cite{DiLisio2023, Svoboda2023, Cangialosi2023}, experimental probes of time-reversibility and validity boundaries of material-time scaling during aging~\cite{Bohmer2024, Henot2024}, real-space models of rejuvenation and memory effects~\cite{Rahman2024}, and analyses of enthalpy fluctuations and fictive-temperature dynamics~\cite{Bagchi2026}.
Nevertheless, recovering Tool's fictive temperature $T_f$ directly from evolving spatial correlations to generate non-equilibrium Kauzmann diagrams, relaxational heat capacities, and the critical scaling of $T_g(q)$ under continuous cooling remained an open challenge.

In this work, we address this problem by combining the microscopic inner-clock framework~\cite{Peredo2022} with the piece-wise NESCGLE formalism developed by Lira-Escobedo \textit{et al.}~\cite{Johnathan2019}, which enables us to subject a benchmark glass-forming liquid of Weeks--Chandler--Andersen (WCA) soft spheres~\cite{WCA1597, PY1958, VW1972, LuisEnrique2013, Mendoza2022} to continuous cooling ramps across seven orders of magnitude in cooling rate ($q \in [10^{-4}, 10^2]$) and the analytical quasi-static limit ($q \to 0$). 
By benchmarking the evolving pair correlations $S(k; u(t))$ and $g(r; u(t))$ against the equilibrium Ornstein--Zernike (OZE) liquid branch, we extract Tool's fictive temperature $T_f(T(t); q)$ 
and uncover a continuous mode fictive-temperature spectrum $T_f(k; u(t); q)$ governed by de Gennes narrowing. 
From the two-body pair-correlation excess entropy $s_2(u(t); q)/k_B$ and the normalized relaxational heat capacity $\tilde{C}_p(T(t); q) = \mathrm{d}T_f/\mathrm{d}T(t)$ along continuous cooling ramps, we construct first-principles Kauzmann diagrams that dynamically avert the entropy crisis above $T_K$~\cite{Kauzmann1948, Debenedetti2001, Mauro2009, McKenna2020, Lunkenheimer2023}. 
Finally, we compare four operational calorimetric definitions of $T_g(q)$ upon cooling~\cite{Moynihan1976, Yue2004, Weyer2005, Cangialosi2023} and derive the critical power-law scaling $T_g(q) - T_c = A\,q^{\psi}$.
Toward the dynamic arrest boundary $T_c > T_K$, governed by the SCGLE critical exponent $\psi = 1/(\gamma + 1) \approx 0.358$, where $\gamma \approx 1.80$ is the equilibrium dynamic critical exponent governing the power-law divergence of the structural relaxation time upon approaching $T_c$.

\section{Theoretical Framework: NESCGLE Theory and the Microscopic Inner Clock}\label{sec:TheoreticalFramework}

\subsection{Onsager--Machlup Canonical Relaxation, the Microscopic Inner Clock $u(t)$, and Piece-Wise Thermal Protocols}\label{subsec:NESCGLE_InnerClock}
Within the non-equilibrium Onsager--Machlup theory of thermal fluctuations~\cite{Onsager1931_1, Onsager1931_2, Onsager1953, Machlup1953, Keizer1987, Pedro2010, Pedro2010_2, Peredo2024, MedinaNoyola2026AnnRev}, Peredo-Ortiz \textit{et al.}~\cite{Peredo2022} showed that when a sequence of elementary instantaneous thermal steps is parameterized by an internal material clock---defined as the time integral of the state-dependent relaxation rate---the resulting structural relaxation automatically assumes the linear-convolution structure of the TNM formalism.
This mapping arises naturally when the dynamics is parameterized by an internal material clock defined as the time integral of the state-dependent relaxation rate. 
For a monocomponent fluid of $N$ spherical particles in volume $V$ (number density $\overline{n} = N/V$, diameter $\sigma$, and volume fraction $\phi = \pi\overline{n}\sigma^3/6$), the fundamental collective structural variable is the non-equilibrium static structure factor $S(k; t) \equiv N^{-1}\langle \delta n(\bm{k}, t)\delta n(-\bm{k}, t)\rangle$~\cite{Pedro2010, Pedro2010_2, Peredo2024, MedinaNoyola2026AnnRev}.

Under the Non-Equilibrium Self-Consistent Generalized Langevin Equation (NESCGLE) theory~\cite{Pedro2010, Pedro2010_2, LuisEnrique2013, Paty2017, Mendoza2022}, each wavenumber $k$ of $S(k;t)$ at reservoir temperature $T$ evolves according to 
\begin{equation}\label{eq:NESCGLE_dSdt_cont}
\frac{\partial S(k; t)}{\partial t} = -2 k^2 D^0(T)\,b(t)\,\overline{n}\mathcal{E}(k; T) \left\{ S(k; t) - \left[\overline{n}\mathcal{E}(k; T)\right]^{-1} \right\},
\end{equation}
where $D^0(T)$ is the short-time self-diffusion coefficient, $b(t) \equiv D_L(t)/D^0(T)$ is the dimensionless instantaneous collective particle mobility~\cite{Pedro2010_2, LuisEnrique2013, Mendoza2022}, and $\mathcal{E}(k; T)$ is the Fourier-transformed \emph{thermodynamic stability matrix} 
~\cite{Pedro2010, Pedro2010_2, Peredo2022, Peredo2024}, defined as the second functional derivative of the Helmholtz free energy density functional evaluated at the uniform number density $\overline{n}$ and temperature $T$. 
Only in true thermodynamic equilibrium (where $\partial S/\partial t = 0$ with $b(t) > 0$) does the stationary solution of Eq.~\eqref{eq:NESCGLE_dSdt_cont} satisfy the Ornstein--Zernike thermodynamic relation $S^{\text{eq}}(k; T) = [\overline{n}\mathcal{E}(k; T)]^{-1}$.
Out of equilibrium, and inside the dynamically arrested regime ($b(t) \to 0$) $\mathcal{E}(k; T)$ remains the fundamental thermodynamic stability matrix driving the structural relaxation.

Equation~\eqref{eq:NESCGLE_dSdt_cont} factorizes the structural relaxation rate into the mode-independent mobility $b(t)$ and the mode-dependent rate $\Gamma(k; T)$:
\begin{equation}\label{eq:NESCGLE_mode_rate}
\tau_{\alpha S}^{-1}(k; t) = b(t)\,\Gamma(k; T), \qquad \Gamma(k; T) \equiv 2 k^2 D^0(T)\,\overline{n}\mathcal{E}(k; T).
\end{equation}
Because $b(t)$ is independent of $k$, introducing the \emph{microscopic inner clock} (accumulated collective mobility)~\cite{Pedro2010_2, Peredo2022, JoaquinJaime2025PoF}
\begin{equation}\label{eq:u_def}
u(t) \equiv \int_0^t b(t')\,\mathrm{d}t', \qquad \mathrm{d}u = b(t)\,\mathrm{d}t,
\end{equation}
linearizes Eq.~\eqref{eq:NESCGLE_dSdt_cont} in $u$ for all wavenumbers $k$ simultaneously:
\begin{equation}\label{eq:dS_du}
\frac{\partial S(k; u)}{\partial u} + \Gamma(k; T)\,S(k; u) = \Gamma(k; T)\,\left[\overline{n}\mathcal{E}(k; T)\right]^{-1}.
\end{equation}
Under spatially homogeneous isochoric conditions ($\overline{n}(\bm{r}, t) = \overline{n}$) and for an \emph{instantaneous quench} to a stationary temperature $T$ (where $\mathcal{E}(k; T)$ and $\Gamma(k; T)$ remain constant for $u > 0$), Eq.~\eqref{eq:dS_du} admits the exact analytical solution~\cite{Pedro2010, Pedro2010_2, LuisEnrique2013, Peredo2022}
\begin{equation}\label{eq:S_analytical_single}
S(k; u) = \left[\overline{n}\mathcal{E}(k; T)\right]^{-1} + \left\{ S(k; u=0) - \left[\overline{n}\mathcal{E}(k; T)\right]^{-1} \right\} \exp\!\left[-\Gamma(k; T)\,u\right].
\end{equation}

To treat finite cooling rates $q \equiv |\mathrm{d}T/\mathrm{d}t|$, Lira-Escobedo \textit{et al.}~\cite{Johnathan2019} represented a continuous thermal protocol $T(t)$ between $T_0$ and $T_M$ as a \emph{concatenation of $M$ isothermal steps} $t_{l-1} \le t < t_l$ ($l = 1, \dots, M$) at discrete temperatures $T_l = T_0 - l\Delta T$ of duration $\Delta t = \Delta T/q$. 
Within each step $l$, $\mathcal{E}(k; T_l)$ is constant; applying Eq.~\eqref{eq:S_analytical_single} with initial condition $S_{l-1}(k; u_{l-1})$ and step inner-clock increment
\begin{equation}\label{eq:step_u_def}
\Delta u_l \equiv u(t_l) - u(t_{l-1}) = \int_{t_{l-1}}^{t_l} b_l(t')\,\mathrm{d}t'
\end{equation}
yields the exact stage-to-stage recurrence relation ($l = 1, 2, \dots, M$)~\cite{Johnathan2019, Peredo2022, JoaquinJaime2025PoF}:
\begin{equation}\label{eq:S_recurrence}
S_l(k; u_l) = \left[\overline{n}\mathcal{E}(k; T_l)\right]^{-1} + \left\{ S_{l-1}(k; u_{l-1}) - \left[\overline{n}\mathcal{E}(k; T_l)\right]^{-1} \right\} \exp\!\left[-\Gamma_l(k)\,\Delta u_l\right],
\end{equation}
where $\Gamma_l(k) \equiv \Gamma(k; T_l)$. 
Unrolling Eq.~\eqref{eq:S_recurrence} recursively from an initial equilibrium liquid state $S_0(k) = S^{\text{eq}}(k; T_0) = [\overline{n}\mathcal{E}(k; T_0)]^{-1}$ produces the closed-form $M$-step piece-wise superposition formula:
\begin{equation}\label{eq:S_TNM_exact}
S_M(k; u_M) = \left[\overline{n}\mathcal{E}(k; T_M)\right]^{-1} - \sum_{l=1}^{M} \left\{ \left[\overline{n}\mathcal{E}(k; T_l)\right]^{-1} - \left[\overline{n}\mathcal{E}(k; T_{l-1})\right]^{-1} \right\} \prod_{j=l}^{M} \exp\!\left[-\Gamma_j(k)\,\Delta u_j\right].
\end{equation}

Equation~\eqref{eq:S_TNM_exact} bridges two key physical limits. In the instantaneous-quench limit ($q \to \infty$, so $\Delta u_j \to 0$ during the ramp), the telescoping sum collapses to $S_M(k) = [\overline{n}\mathcal{E}(k; T_0)]^{-1}$ and subsequent isothermal relaxation at $T_{\text{end}} = T_M$ reduces identically to the single-jump solution~\eqref{eq:S_analytical_single}~\cite{Pedro2010, Pedro2010_2, LuisEnrique2013, Mendoza2022}. 
Conversely, in the continuous-ramp limit ($M \to \infty$, $\Delta t \to 0$ at finite $q$), the Riemann sum in Eq.~\eqref{eq:S_TNM_exact} converges to the continuous Volterra hereditary integral of the TNM formalism~\cite{Tool1946, Narayanaswamy1971, Moynihan1976, Peredo2022}:
\begin{equation}\label{eq:S_Volterra_TNM}
S(k; u(t)) = \left[\overline{n}\mathcal{E}(k; T(t))\right]^{-1} - \int_0^t \mathrm{d}t' \left\{ \frac{\partial \left[\overline{n}\mathcal{E}(k; T(t'))\right]^{-1}}{\partial t'} \right\} \exp\!\left( -\int_{u(t')}^{u(t)} \Gamma(k; T(u''))\,\mathrm{d}u'' \right).
\end{equation}

Physically, each increment $\Delta u_l$ in Eq.~\eqref{eq:step_u_def} measures the ratio of the residence time $\Delta t = \Delta T/q$ to the structural relaxation time $\tau_\alpha(t) \propto 1/b(t)$ (the inverse local Deborah number $\mathrm{De}_l^{-1}$). 
At high temperatures where $\Delta u_l \gg 1$ ($\mathrm{De}_l \ll 1$), $\exp[-\Gamma_l(k)\Delta u_l] \to 0$ and the liquid equilibrates to $S^{\text{eq}}(k; T_l) = [\overline{n}\mathcal{E}(k; T_l)]^{-1}$ at each step. 
Upon deeper cooling where $b(t) \to 0$, $\Delta u_l \to 0$ ($\mathrm{De}_l \gg 1$) and the material clock arrests at a finite asymptotic value $u^a(q) < \infty$. 
Evaluating $b(t) = b[S(k; u(t))]$ and the squared localization length $\gamma(u)$ along the trajectory is accomplished via the NESCGLE Self-Consistent Scheme in Appendix~\ref{app:SCGLE_Closure}.

\subsection{The WCA Soft-Sphere Glass Former and Evolution of the Inner Clock $u(t)$}\label{subsec:WCA_Model}

To illustrate our thermodynamic framework on a concrete microscopic model, we consider a monodisperse fluid of particles interacting via the purely repulsive Weeks--Chandler--Andersen (WCA) pair potential with softness exponent $\nu = 6$:
\begin{equation}\label{eq:WCA_potential}
\beta u_{\text{WCA}}(r) = \begin{cases}
\dfrac{1}{T^*}\left[ \left(\dfrac{\sigma}{r}\right)^{2\nu} - 2\left(\dfrac{\sigma}{r}\right)^{\nu} + 1 \right], & r \le \sigma, \\[10pt]
0, & r > \sigma,
\end{cases}
\end{equation}
where $T^* \equiv k_B T/\epsilon = (\beta\epsilon)^{-1}$ denotes the dimensionless reduced temperature  and $\epsilon$ is the repulsive energy scale. 

Throughout this work, length is measured in units of $\sigma$, time $t$ in units of $t_0 \equiv \sigma^2/D^0$, and the cooling rate $q \equiv |\mathrm{d}T^*/\mathrm{d}t|$ in units of $\epsilon/(k_B t_0)$; in dense colloidal suspensions, short-time hydrodynamic interactions simply renormalize $D^0$ to the cage diffusivity $D_S(\phi)$ without altering $\mathcal{E}(k; T)$ or the excluded-volume arrest scenario~\cite{MedinaNoyola1988, LopezFlores2012}. 
Adopting natural energy units $\epsilon/k_B \equiv 1$ renders $T$ and $T^*$ numerically identical, so all temperatures reported below ($T(t)$, $T_f(t)$, $T_c$, $T_K$, and $T_g(q)$) directly denote reduced temperatures $T^*$.

This WCA fluid provides a thoroughly validated benchmark whose equilibrium scaling~\cite{Pedro2011}, quench aging~\cite{LuisEnrique2013, Paty2017, Mendoza2022}, piece-wise cooling~\cite{Johnathan2019}, inner-clock kinetics~\cite{Peredo2022}, and linear viscoelasticity~\cite{Peredo2025} are well established.
Following Refs.~\cite{LuisEnrique2013, Mendoza2022, Peredo2025}, the inverse stability matrix $[\overline{n}\mathcal{E}(k; \phi, T)]^{-1}$ is obtained via the WCA blip perturbation theory~\cite{WCA1597, HansenMcDonald1990}, mapping the soft-core fluid onto a Percus--Yevick~\cite{PY1958} and Verlet--Weis~\cite{VW1972} hard-sphere reference fluid at the effective packing fraction
\begin{equation}\label{eq:blip_mapping}
\phi_{\text{eff}}(\phi, T) = \phi\,\lambda_{\text{blip}}^3(T), \qquad \lambda_{\text{blip}}^3(T) = 1 - 3\int_0^1 x^2 \exp\!\left[-\beta u_{\text{WCA}}(x\sigma)\right]\mathrm{d}x.
\end{equation}

\begin{figure}[t!]
    \centering
    \includegraphics[width= 0.45\textwidth]{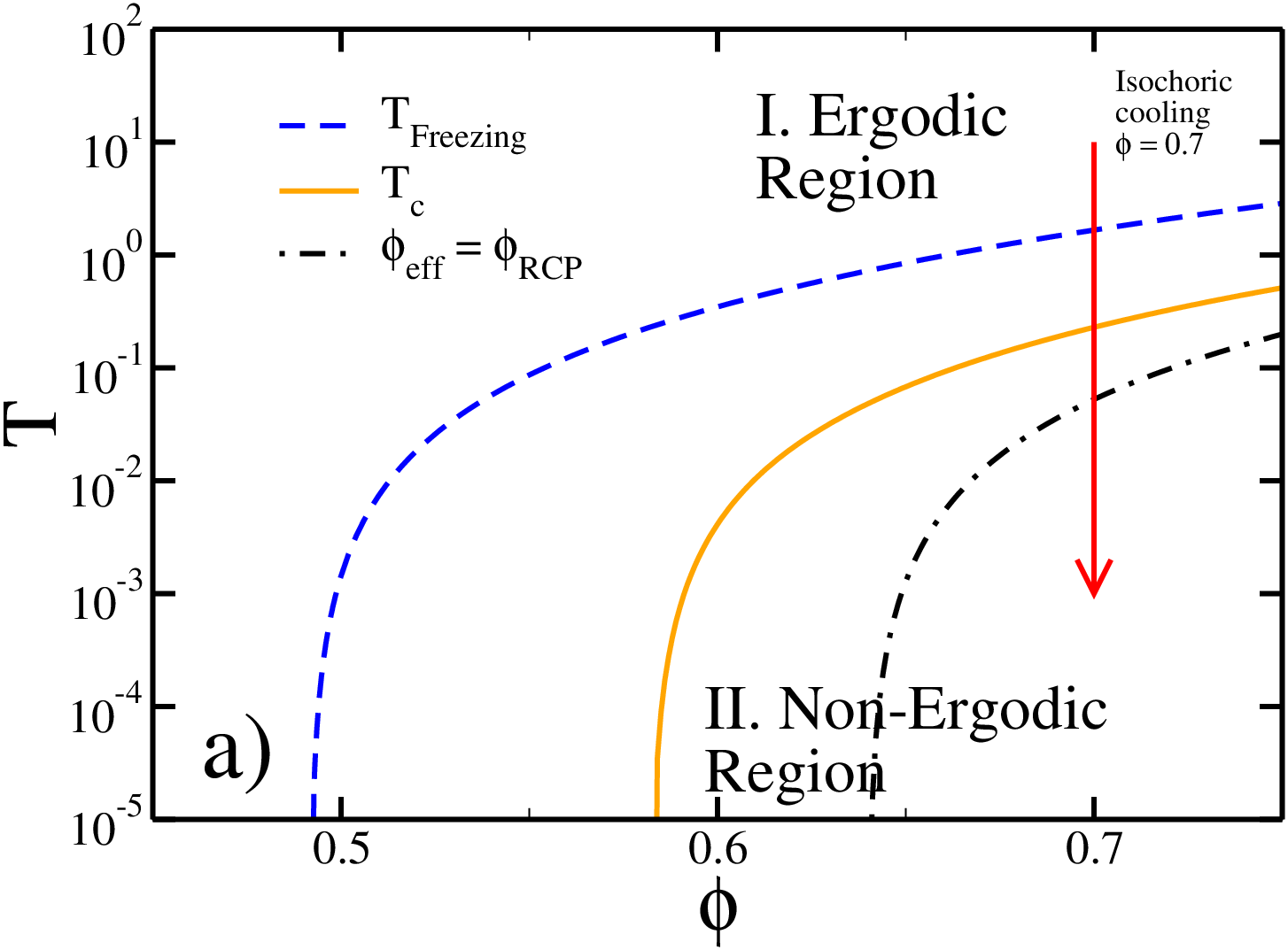}
    \includegraphics[width= 0.45\textwidth]{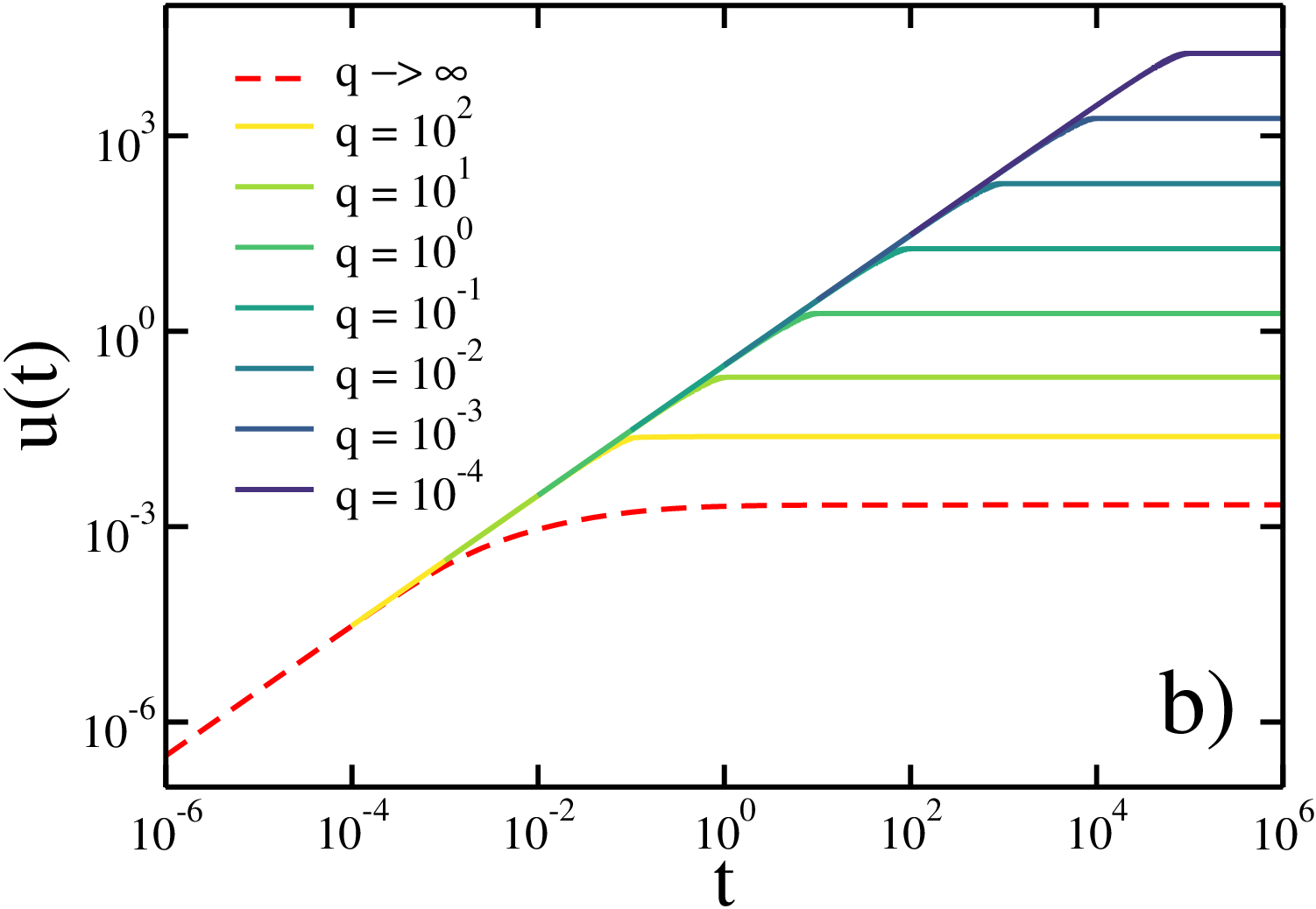}
    \caption{
    \textbf{(a)} Dynamic arrest diagram in the $(\phi, T)$ plane ($T \equiv T^* \equiv k_B T/\epsilon$) showing the equilibrium freezing boundary $T_{\text{Freezing}}(\phi)$ ($\phi_{\text{eff}} \approx 0.494$, blue dashed curve), the ideal dynamic arrest boundary $T_c(\phi)$ obtained from Eq.~\eqref{eq:gamma_fixed_point} ($\phi_{\text{eff}}^c \approx 0.582$, solid orange curve), the extrapolated Kauzmann / Random Close Packing boundary $T_K(\phi)$ ($\phi_{\text{eff}} = \phi_{\text{RCP}} = 0.64$, black dash-dotted curve), and the primary isochoric cooling path at $\phi = 0.70$ (vertical red arrow). \textbf{(b)} Evolution of the accumulated collective mobility (TNM material time) $u(t) = \int_0^t b(t')\,\mathrm{d}t'$ as a function of chronological laboratory time $t$ along the $\phi = 0.70$ isochore for cooling from $T_0 = 10.0$ to $T_{\text{end}} = 10^{-3}$ across cooling rates $q \in \{\infty, 10^2, 10^1, 10^0, 10^{-1}, 10^{-2}, 10^{-3}, 10^{-4}\}$.}
    \label{fig:arrest_wca}
\end{figure}

Evaluating the localization criterion~\eqref{eq:gamma_fixed_point} with the equilibrium input $[\overline{n}\mathcal{E}(k; \phi, T)]^{-1} = S^{\text{eq}}(k; \phi, T)$ yields the ideal arrest line $T_c(\phi)$ in Fig.~\ref{fig:arrest_wca}(a) along the contour $\phi_{\text{eff}}(\phi, T_c) = \phi_c^{\text{HS}} \approx 0.582$~\cite{Juarez2007, Pedro2011, LuisEnrique2013, Mendoza2022}. 
Here we follow the representative overpacked isochore $\phi = 0.70$ [vertical red arrow in Fig.~\ref{fig:arrest_wca}(a)]: cooling from $T_0 = 10.0$ crosses the freezing line $T_{\text{Freezing}}(\phi)$ ($\phi_{\text{eff}} \approx 0.494$, blue dashed line), traverses the ideal arrest boundary at $T_c = 0.2287$ ($1/T_c = 4.372$, solid orange line), and approaches the Kauzmann/RCP limit at $T_K = 0.0525$ ($\phi_{\text{eff}}^K = \phi_{\text{RCP}} \approx 0.64$, black dash-dotted line). 
Because the NESCGLE equations enforce macroscopic homogeneity and isotropy ($\overline{n}(\bm{r}, t) = \overline{n}$), crystal nucleation is excluded by construction, cleanly isolating the amorphous branch as in standard hard-sphere mode-coupling theory~\cite{Goetze2009} and matching Brownian dynamics benchmarks with $\sim 10\%$ polydispersity ~\cite{LuisEnrique2013, Paty2017, Mendoza2022}. 
For any finite cooling rate $q > 0$, crossing below $T_c$ drives $b(t) \to 0$ and arrests the inner clock $u(t)$ at a finite plateau $u^a(q) \equiv \lim_{t\to\infty} u(t) < \infty$ [Fig.~\ref{fig:arrest_wca}(b)]. 
Slower cooling rates remain longer in the mobile regime, accumulating larger frozen material times $u^a(q)$ that lock $S(k; u(t))$ into a $q$-dependent glass state $S(k; u^a(q))$ and kinetically avert the Kauzmann catastrophe at $T_K$.
\section{Recovering Tool's Fictive Temperature from Non-Equilibrium Spatial Correlations}\label{sec:FictiveTemperature}

\subsection{Non-Equilibrium Structural Evolution Along Continuous Cooling Ramps}\label{subsec:Sk_and_gr}

Having established how the cooling rate $q$ controls the arrest of the microscopic inner clock $u(t)$, we now examine the continuous evolution of the non-equilibrium liquid and glass microstructure along a single deep cooling ramp from an initial equilibrium fluid state at $T_i = 10.0$ down to the bath temperature $T_{\text{end}} = 10^{-3}$ ($M = 1000$ stages of width $\Delta T \approx 0.01$ at $\phi = 0.70$) across seven orders of magnitude in cooling rate, $q \in \{10^2, 10^1, 10^0, 10^{-1}, 10^{-2}, 10^{-3}, 10^{-4}\}$, together with the ideal quasi-static cooling limit $q \to 0$. 
Rather than restricting our analysis to asymptotic long-time endpoints following separate instantaneous quenches, we track the structural evolution continuously along the thermal ramp as a function of the bath temperature,
\begin{equation}\label{eq:T_inst_def}
T(t) = T_i - q t,
\end{equation}
and the accumulated material time $u(t)$. 
Specifically, at each stage of the cooling process, the instantaneous static structure factor $S(k; u(t))$ and radial distribution function $g(r; u(t))$ are benchmarked against the reference equilibrium branch, $S^{\text{eq}}(k; T) = [\overline{n}\mathcal{E}(k; T)]^{-1}$ and $g^{\text{eq}}(r; T)$, obtained from the Ornstein--Zernike relation closed via Eq.~\eqref{eq:blip_mapping}.
Within this framework, the quasi-static cooling limit $q \to 0$ admits an exact analytical characterization: since $\mathrm{d}u/\mathrm{d}T = -b(t)/q$, at every bath temperature $T(t) > T_c$ where the equilibrium mobility remains strictly positive ($b_{\text{eq}}(T) > 0$), taking $q \to 0^+$ yields an infinite material-time increment $\Delta u_l \to \infty$ in Eq.~\eqref{eq:S_recurrence}, so that $\exp[-\Gamma_l(k)\Delta u_l] \to 0$ and the system tracks the exact Ornstein--Zernike equilibrium branch $S^{\text{OZ}}(k; T) \equiv S^{\text{eq}}(k; T)$ down to $T_c = 0.2287$. 
Conversely, as soon as the bath temperature crosses below the ideal dynamic arrest boundary ($T(t) \le T_c$), the collective mobility vanishes identically ($b_{\text{eq}}(T \le T_c) = 0$), halting the inner clock ($\Delta u_l = 0$) and permanently arresting the microstructure in the critical configuration $S^{\text{eq}}(k; T_c)$ for all $T(t) \le T_c$.

\begin{figure}[t!]
    \centering
    \includegraphics[width=0.32\textwidth]{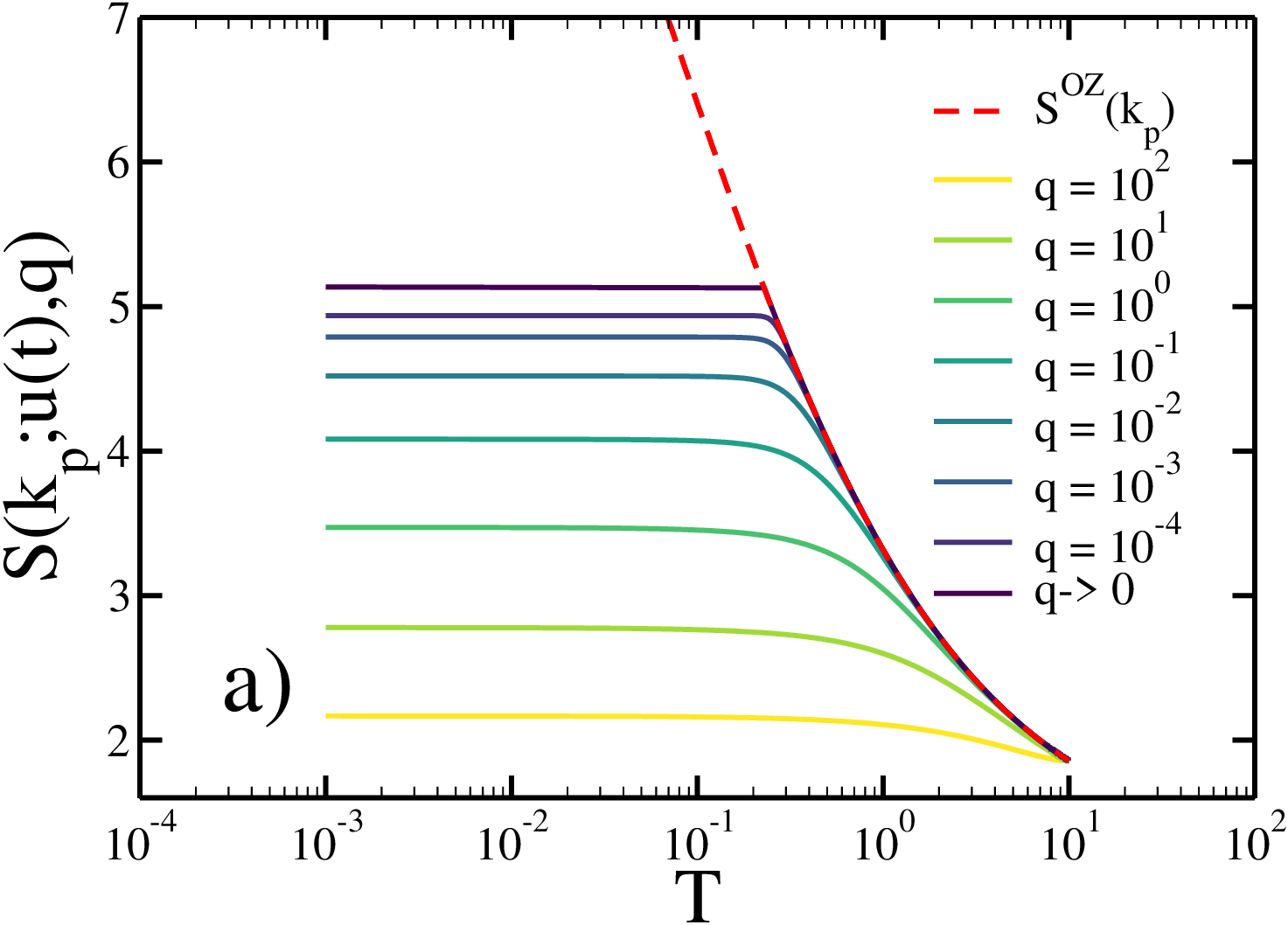}
    \includegraphics[width=0.32\textwidth]{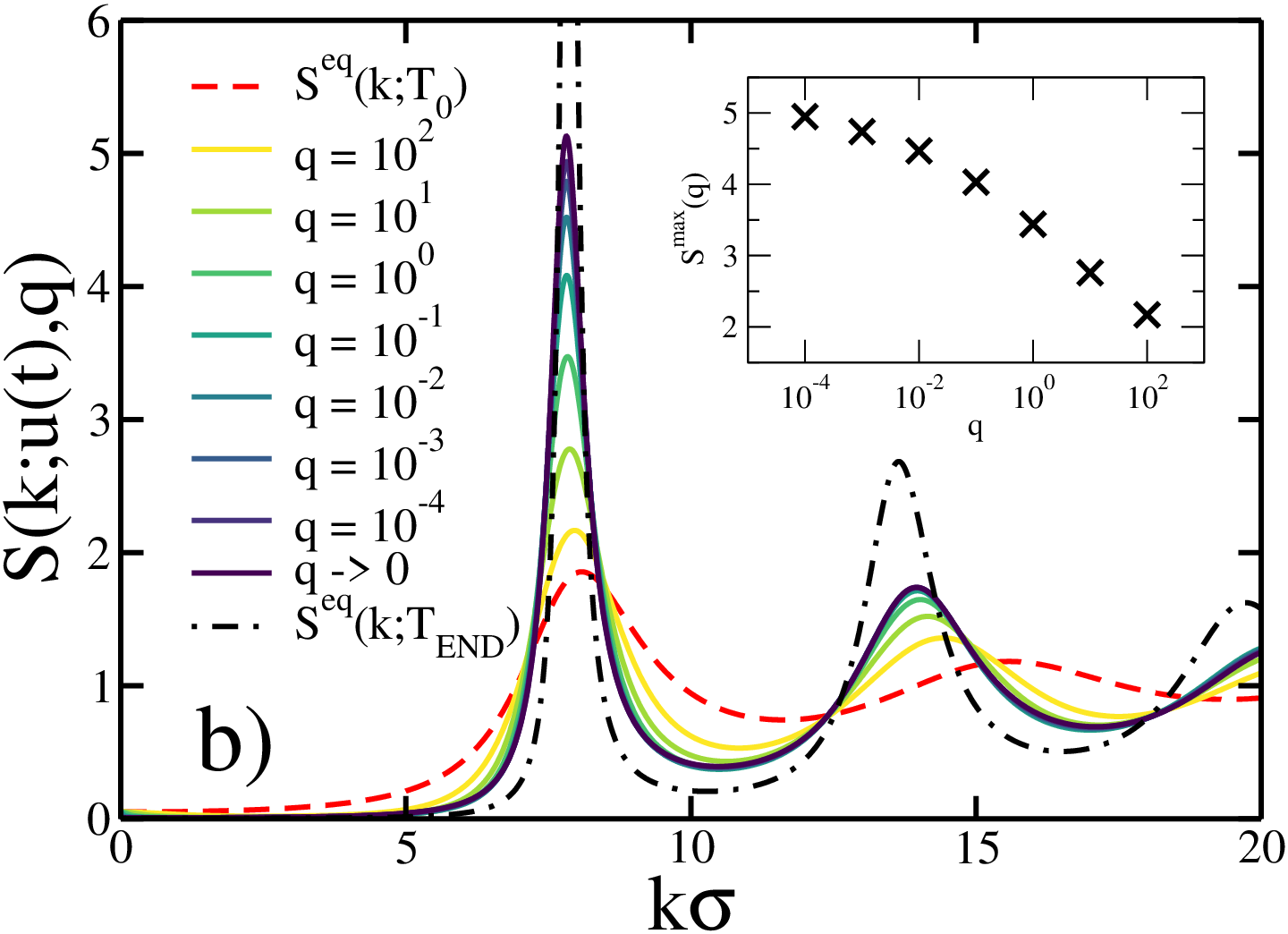}
    \includegraphics[width=0.32\textwidth]{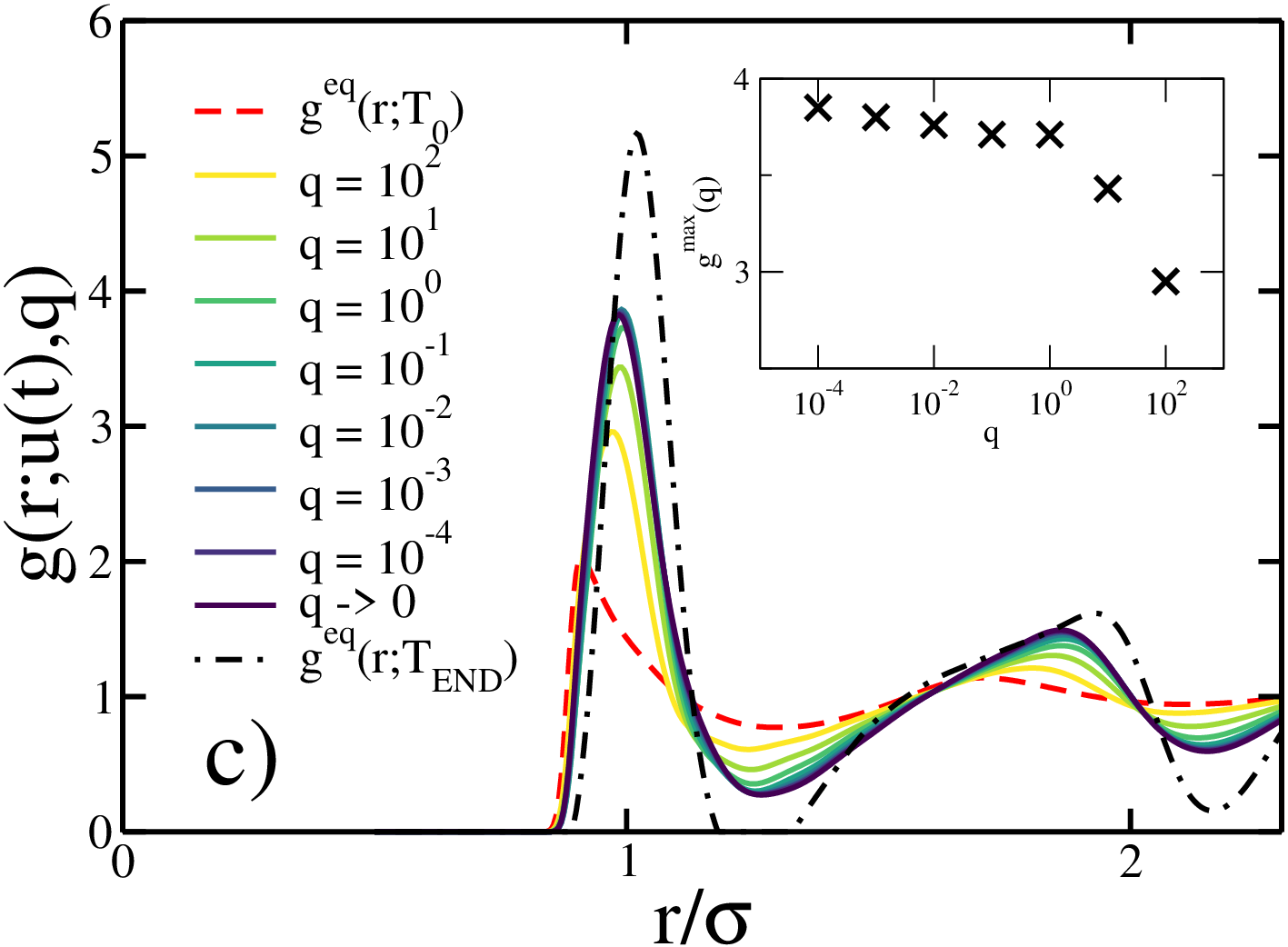}
    \caption{
    \textbf{(a)} Evolution of the main diffraction peak height $S(k_p; u(t), q)$ along the cooling ramp as a function of the instantaneous bath temperature $T(t) = T_i - q t$ across $q \in [10^{-4}, 10^2]$ and the quasi-static limit $q \to 0$ (which arrests sharply at $T_c = 0.2287$), compared against the Ornstein--Zernike equilibrium branch $S^{\text{OZ}}(k_p; T)$ (red dashed curve). \textbf{(b)} Instantaneous static structure factor $S(k; u(t); q)$ vs dimensionless wavenumber $k\sigma$ at the end of the cooling ramp ($T(t) = T_{\text{END}} = 10^{-3}$) for $q \in [10^{-4}, 10^2]$ and $q \to 0$, benchmarked against the equilibrium profiles at the initial temperature $S^{\text{eq}}(k; T_0 = 10.0)$ (red dashed curve) and at the target temperature $S^{\text{eq}}(k; T_{\text{END}} = 10^{-3})$ (black dash-dotted curve); the inset displays the arrested primary peak height $S^{\max}(q)$ vs $q$. \textbf{(c)} Corresponding instantaneous radial distribution function $g(r; u(t); q)$ at $T(t) = T_{\text{END}} = 10^{-3}$ vs $r/\sigma$ compared against $g^{\text{eq}}(r; T_0 = 10.0)$ (red dashed curve) and $g^{\text{eq}}(r; T_{\text{END}} = 10^{-3})$ (black dash-dotted curve), with the inset showing the arrested nearest-neighbor contact peak $g^{\max}(q)$ vs $q$.}
    \label{fig:structure_Sk_gr}
\end{figure}

The physical picture that emerges from this continuous cooling protocol is illustrated in Fig.~\ref{fig:structure_Sk_gr}. 
Figure~\ref{fig:structure_Sk_gr}(a) tracks the continuous thermal evolution of the main diffraction peak height $S(k_p; u(t), q)$ as a function of the instantaneous bath temperature $T(t)$ across all cooling rates $q \in [10^{-4}, 10^2]$ and the quasi-static limit $q \to 0$, compared directly against the Ornstein--Zernike equilibrium solution $S^{\text{OZ}}(k_p; T)$ (red dashed curve). 
At high temperatures, all cooling trajectories follow $S^{\text{OZ}}(k_p; T)$ as the effective hard-sphere diameter $\lambda(\phi, T)$ expands; upon deeper cooling, however, the progressive slowing down of the inner clock $u(t)$ causes each non-equilibrium trajectory to peel off from the equilibrium curve and saturate onto a rate-ordered horizontal plateau bounded between the fastest quench ($q = 10^2$) and the quasi-static limit ($q \to 0$, which tracks $S^{\text{OZ}}(k_p; T)$ down to $T_c = 0.2287$ and arrests sharply at $S_{\max}^{\text{eq}}(T_c) \approx 5.25$), whereas the hypothetical equilibrium branch $S^{\text{OZ}}(k_p; T)$ continues rising steeply toward low temperatures. 
The resulting arrested microstructures at the end of the cooling ramp ($T(t) = T_{\text{END}} = 10^{-3}$) are resolved across reciprocal and real space in Figs.~\ref{fig:structure_Sk_gr}(b) and \ref{fig:structure_Sk_gr}(c). 
At the beginning of the ramp ($T_i = 10.0$), thermal agitation easily overcomes the soft repulsive core of the WCA potential ($\phi_{\text{eff}}(T_i) = 0.3808$), and the fluid displays the broad, weakly correlated structure depicted by the red dashed reference curves in Figs.~\ref{fig:structure_Sk_gr}(b) and \ref{fig:structure_Sk_gr}(c).
As the thermal bath cools toward the final bath temperature $T_{\text{end}} = 10^{-3}$, the effective hard-sphere diameter $\lambda(\phi, T)$ expands and the system builds up increasingly sharp spatial correlations. 
If the liquid were able to maintain thermodynamic equilibrium all the way down to the target temperature $T_{\text{end}} = 10^{-3}$, the Ornstein--Zernike (OZE) equation would predict the hyper-structured profiles shown by the black dash-dotted curves in Figs.~\ref{fig:structure_Sk_gr}(b) and \ref{fig:structure_Sk_gr}(c), characterized by a towering primary diffraction peak ($S_{\max}^{\text{eq}}(10^{-3}) \approx 11.78$), an intense nearest-neighbor contact peak ($g_1^{\text{eq}}(10^{-3}) \approx 5.15$), a deeply depleted first interstitial minimum, and a sharply resolved second-shell doublet. 
In stark contrast, the non-equilibrium NESCGLE prediction reveals that the cooling liquid never reaches this hypothetical $T_{\text{end}} = 10^{-3}$ equilibrium state; instead, it vitrifies into a family of arrested glassy profiles strictly ordered by the cooling rate $q$. 
Under an ultra-fast quench ($q = 10^2$), the inner clock $u(t)$ freezes rapidly ($u(t_{\text{end}}) \approx 0.0228$), trapping a weakly correlated configuration 
whose main diffraction peak $S(k_p; u(t); q)$ in Fig.~\ref{fig:structure_Sk_gr}(b) and radial distribution $g(r; u(t); q)$ in Fig.~\ref{fig:structure_Sk_gr}(c) remain close to those of the high-temperature fluid. 
As the cooling rate is progressively slowed across seven orders of magnitude down to $q = 10^{-4}$ and toward the quasi-static limit $q \to 0$, the main cage peak around $k_p\sigma \approx 7.1\text{--}7.8$ grows and sharpens 
while long-wavelength density fluctuations ($k \to 0$) are systematically suppressed.

In real space [Fig.~\ref{fig:structure_Sk_gr}(c)], this kinetic hierarchy translates into a progressively more ordered local environment: the nearest-neighbor contact shell around $r/\sigma \approx 1$ intensifies, the first minimum deepens, and the second coordination shell ($r/\sigma \in [1.75, 2.15]$) develops the characteristic shoulder and splitting associated with dense amorphous packing~\cite{Vollmayr1996, Truskett2000, Li2011}. 
The monotonic growth of both the arrested primary diffraction peak $S^{\max}(q)$ and the arrested nearest-neighbor contact peak $g^{\max}(q)$ upon decreasing $q$ is summarized in the insets of Figs.~\ref{fig:structure_Sk_gr}(b) and \ref{fig:structure_Sk_gr}(c).
Most notably, comparing the non-equilibrium family against the equilibrium benchmarks in Figs.~\ref{fig:structure_Sk_gr}(a)--\ref{fig:structure_Sk_gr}(c) exposes a striking result: whereas the fastest quench ($q = 10^2$) is arrested in a high-temperature fluid-like configuration, even when the cooling rate is reduced by seven orders of magnitude down to $q = 10^{-4}$ or taken to the quasi-static limit $q \to 0$, the slowest cooling process remains far below the $T_{\text{end}} = 10^{-3}$ OZE prediction and becomes trapped directly in the structural configuration corresponding to the critical arrest temperature $T_c = 0.2287$ [the $q \to 0$ limit curves $S^{\text{eq}}(k; T_c)$ and $g^{\text{eq}}(r; T_c)$]. 

\subsection{Operational Extraction of Tool's Fictive Temperature $T_f(T(t); q)$ and Wavenumber-Resolved Mode Spectrum}\label{subsec:ToolExtraction}

\begin{figure}[t!]
    \centering
    \includegraphics[width=0.32\textwidth]{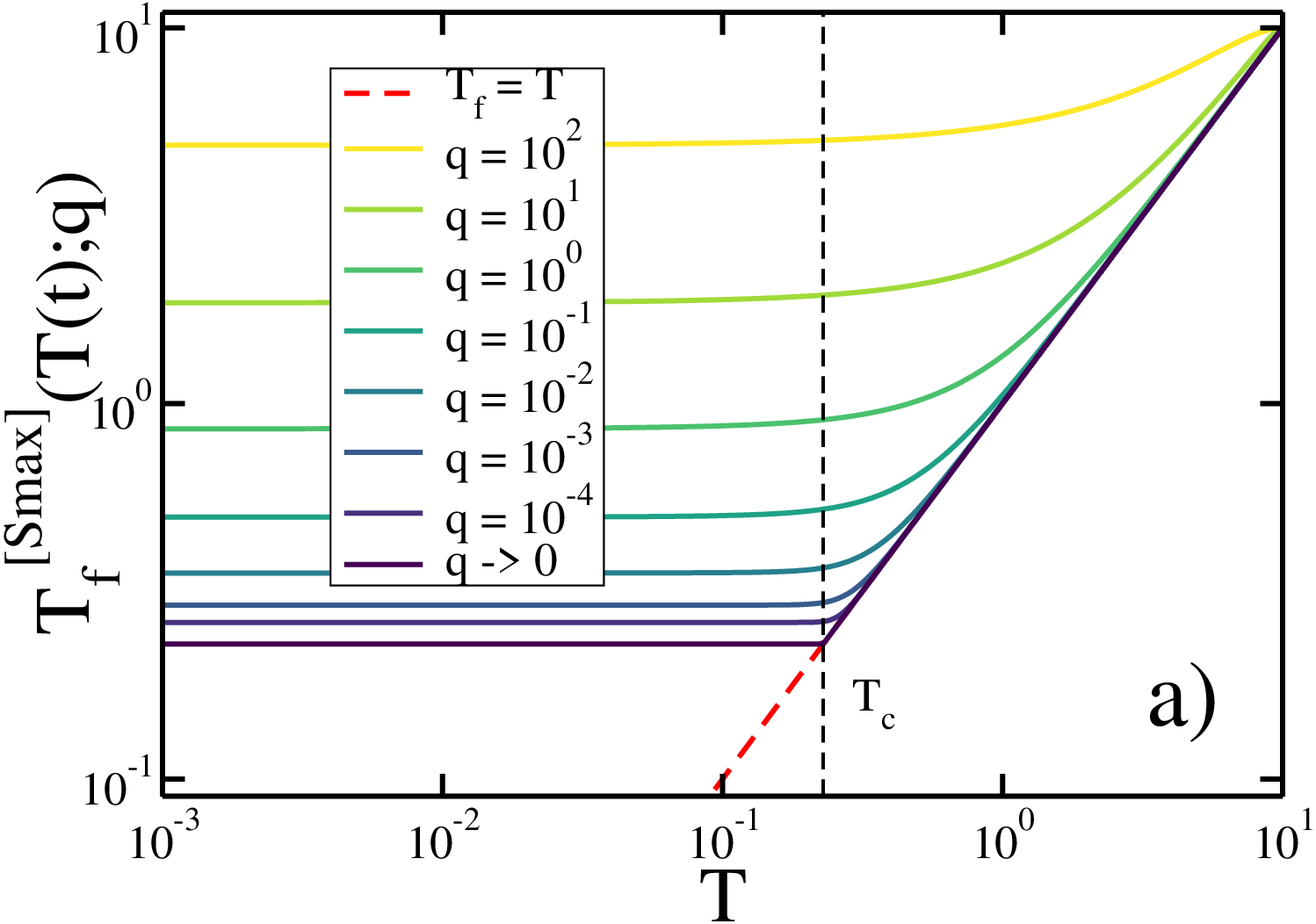}
    \includegraphics[width=0.32\textwidth]{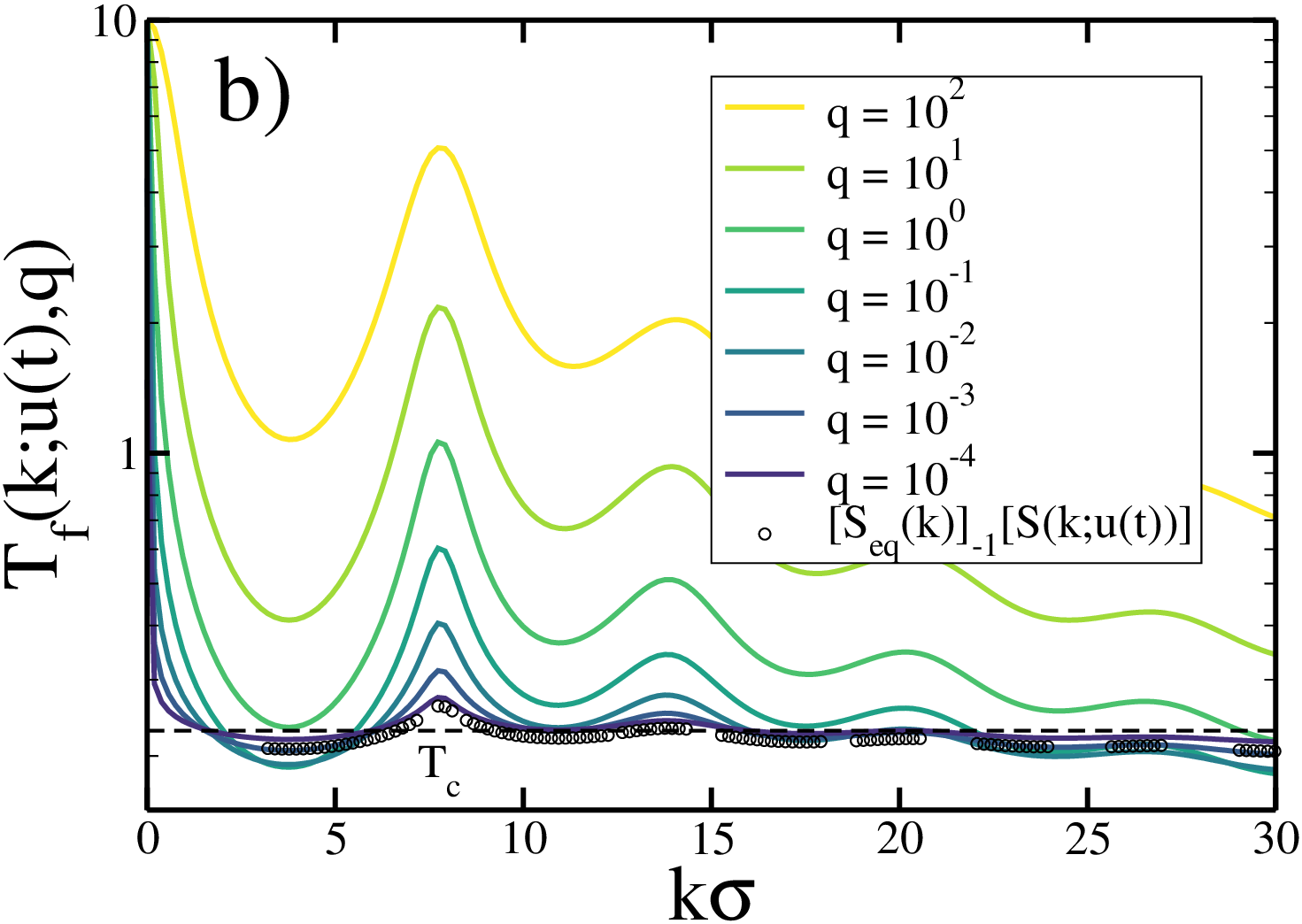}
    \includegraphics[width=0.32\textwidth]{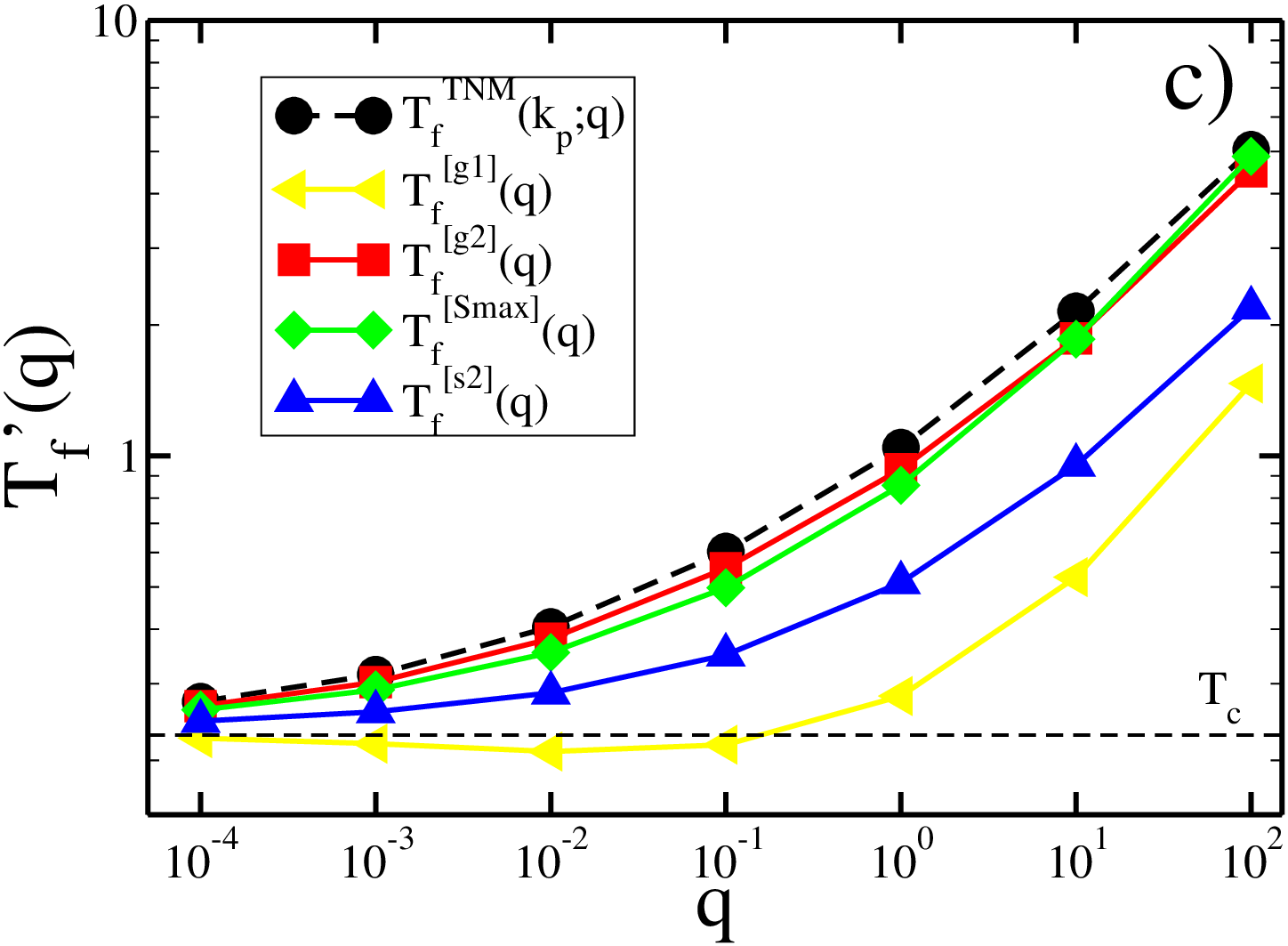}
    \caption{
    \textbf{(a)} Instantaneous structural fictive temperature $T_f^{[S_{\max}]}(T(t); q)$ vs instantaneous bath temperature $T(t) = T_i - q t$ ($T_i = 10.0$, $M = 1000$ steps) across seven decades of cooling rate ($q \in [10^{-4}, 10^2]$) and the quasi-static limit ($q \to 0$, which arrests sharply at $T_c = 0.2287$, vertical dashed line), showing exact tracking of the thermodynamic equilibrium line $T_f = T$ (red dashed line) above $T_c = 0.2287$ and smooth arrest onto rate-ordered limiting glass plateaus $T'_f(q)$ decreasing from $4.873$ ($q = 10^2$) to $0.261$ ($q = 10^{-4}$) and $T_c = 0.2287$ ($q \to 0$). \textbf{(b)} Wavenumber-resolved mode fictive temperature spectrum $T_f(k; u(t); q)$ vs dimensionless wavenumber $k\sigma$ at the end of the cooling ramp ($T(t) = 10^{-3}$), comparing the NESCGLE inner-clock mode propagation $T_f^{\text{TNM}}(k; u(t); q)$ [Eq.~\eqref{eq:Tf_k_TNM}, solid curves] with direct mode-by-mode structural inversion $[S^{\text{eq}}(k)]^{-1}(S(k; u(t)))$ (open circles for $q = 10^{-4}$) and the critical arrest reference $T_c = 0.2287$ (horizontal black dashed line, corresponding to the uniform quasi-static limit $q \to 0$), revealing pronounced de Gennes narrowing at the cage peak $k_p\sigma \approx 7.12$. \textbf{(c)} Moynihan plot of the limiting glass fictive temperatures $T'_f(q) \equiv \lim_{T(t)\to 0} T_f(T(t); q)$ vs cooling rate $q \in [10^{-4}, 10^2]$ across all structural observables (see text) relative to the horizontal critical arrest asymptote $T_c = 0.2287$ (black dashed line).}
    \label{fig:fictive_temperature}
\end{figure}

The departure of the non-equilibrium cooling trajectories from the equilibrium OZE curve in Fig.~\ref{fig:structure_Sk_gr}(a) provides a direct and intuitive way to determine Tool's fictive temperature $T_f(T(t); q)$~\cite{Tool1946}. 
By definition, the fictive temperature of an out-of-equilibrium glass-forming liquid at instant $t$ (when the external thermal bath is at temperature $T(t) = T_i - q t$) is the thermodynamic temperature at which the equilibrium liquid would exhibit the exact same structural value $\mathcal{P}$ as the non-equilibrium system possesses at that instant. 
Within our microscopic framework, solving the equilibrium Ornstein--Zernike equation yields a smooth, strictly monotonic equilibrium curve $\mathcal{P}^{\text{eq}}(T)$ for any structural observable [such as the red dashed curve $S^{\text{OZ}}(k_p; T)$ in Fig.~\ref{fig:structure_Sk_gr}(a)]. 
Because $\mathcal{P}^{\text{eq}}(T)$ is monotonic, there is a one-to-one correspondence between the equilibrium value of the observable and the temperature $T$, so $\mathcal{P}^{\text{eq}}(T)$ acts as a thermometer calibrated in equilibrium. 
At any point along the cooling ramp, the NESCGLE equations provide the instantaneous non-equilibrium value $\mathcal{P}(u(t); q)$, and the corresponding fictive temperature $T_f^{[\mathcal{P}]}(T(t); q)$ is obtained simply by matching this non-equilibrium value to the equilibrium calibration curve:
\begin{equation}\label{eq:Tool_inversion}
\mathcal{P}^{\text{eq}}\!\left(T_f^{[\mathcal{P}]}(T(t); q)\right) = \mathcal{P}(u(t); q) \quad \Longleftrightarrow \quad T_f^{[\mathcal{P}]}(T(t); q) = \left(\mathcal{P}^{\text{eq}}\right)^{-1}\!\left(\mathcal{P}(u(t); q)\right).
\end{equation}
Geometrically, Eq.~\eqref{eq:Tool_inversion} corresponds to a straightforward horizontal projection in Fig.~\ref{fig:structure_Sk_gr}(a): 
starting from a point on a non-equilibrium cooling curve $S(k_p; u(t), q)$ at bath temperature $T(t)$, one moves horizontally at constant peak height until intersecting the red dashed Ornstein--Zernike equilibrium curve $S^{\text{OZ}}(k_p; T)$, and reads the fictive temperature $T_f$ directly from the temperature axis of that intersection point. 
At high temperatures, where the cooling trajectory lies on top of the OZE equilibrium curve, this horizontal projection is trivial and gives $T_f(T(t); q) = T(t)$. 
Once the system falls out of equilibrium and $\mathcal{P}(u(t); q)$ freezes onto a horizontal plateau, projecting that constant plateau height horizontally onto the OZE curve yields the constant limiting fictive temperature $T'_f(q) \equiv \lim_{T(t)\to 0} T_f(T(t); q)$ of the glass ---which in the quasi-static limit $q \to 0$ reduces identically to the critical arrest temperature, $T'_f(q \to 0) = T_c = 0.2287$ [Fig.~\ref{fig:fictive_temperature}(a)]. 
To examine how different spatial scales record this thermal arrest, we apply the calibration of Eq.~\eqref{eq:Tool_inversion} simultaneously along the cooling ramp to four fundamental reciprocal- and real-space structural observables computed directly from $S(k; u(t))$ and $g(r; u(t))$:
\begin{itemize}
    \item $T_f^{[S_{\max}]}(T(t); q)$, inverted from the instantaneous primary diffraction peak height $S_{\max}(u(t); q) = S_{\max}^{\text{eq}}(T_f)$;
    \item $T_f^{[g_1]}(T(t); q)$, inverted from the instantaneous nearest-neighbor contact peak $g_1(u(t); q) = g_1^{\text{eq}}(T_f)$;
    \item $T_f^{[g_2]}(T(t); q)$, inverted from the instantaneous second-shell peak height $g_2(u(t); q) = g_2^{\text{eq}}(T_f)$;
    \item $T_f^{[s_2]}(T(t); q)$, inverted from the instantaneous two-body pair-correlation excess entropy $s_2(u(t); q) = s_2^{\text{eq}}(T_f)$~\cite{Baranyai1989, HansenMcDonald1990, Truskett2000, Pond2011}, which accounts for more than $85\%$ of the total excess entropy of dense simple liquids and is defined solely in terms of the instantaneous radial distribution function $g(r; u(t); q)$ as
    \begin{equation}\label{eq:s2_definition}
    \frac{s_2(u(t); q)}{k_B} = -2\pi\overline{n} \int_0^\infty \left[ g(r; u(t); q) \ln g(r; u(t); q) - g(r; u(t); q) + 1 \right] r^2\,\mathrm{d}r.
    \end{equation}
\end{itemize}

In addition to calibrating macroscopic observables via Eq.~\eqref{eq:Tool_inversion}, the NESCGLE step-by-step cooling recurrence~\eqref{eq:S_recurrence}--\eqref{eq:S_Volterra_TNM} shows that at the microscopic level each spatial wavenumber $k$ possesses its own mode fictive temperature $T_f(k; u(t))$. 
In Eq.~\eqref{eq:S_recurrence}, during each elementary cooling step from $T_{l-1}$ to $T_l$, the structure factor of $k$ relaxes toward its new equilibrium target $[\overline{n}\mathcal{E}(k; T_l)]^{-1}$ through the exponential damping factor $\exp[-\Gamma_l(k)\Delta u_l]$. 
Applying this exact same exponential relaxation factor directly to the temperature steps defines the mode-resolved TNM fictive temperature $T_f^{\text{TNM}}(k; u_l)$ along the cooling ramp:
\begin{eqnarray}\label{eq:Tf_k_TNM}
T_f^{\text{TNM}}(k; u_l) &=& T_l + \left[ T_f^{\text{TNM}}(k; u_{l-1}) - T_l \right] \exp\!\left[-\frac{\Delta u_l}{\tau_{\alpha S}(k; T_l)}\right], \nonumber \\
\tau_{\alpha S}(k; T_l) &\equiv& \frac{1}{\Gamma_l(k)} = \frac{\left[\overline{n}\mathcal{E}(k; T_l)\right]^{-1}}{2 k^2 D^0(T_l)}.
\end{eqnarray}
Crucially, $T_f(k; u(t); q)$ is not a modal kinetic temperature---fast momentum and intra-cage vibrations remain thermalized at the uniform bath temperature $T(t)$---but rather Tool's equilibrium calibration temperature~\cite{Tool1946} for density fluctuations of wavelength $2\pi/k$ [Fig.~\ref{fig:fictive_temperature}(b)]. 
Because the material-time relaxation time $\tau_{\alpha S}(k; T_l) \propto [\overline{n}\mathcal{E}(k; T_l)]^{-1}/k^2$ peaks sharply at the main diffraction peak $k_p\sigma \approx 7.12\text{--}7.85$ due to weak thermodynamic restoring forces (de Gennes narrowing~\cite{deGennes1959}), cage-scale and $k \to 0$ modes decouple first upon cooling and freeze at the highest $T_f(k; u(t); q)$. 
Conversely, modes in the troughs of $S(k)$ or at higher shells ($k_2\sigma \approx 14.1$, $k_3\sigma \approx 20.6$) have shorter relaxation times $\tau_{\alpha S}(k; T_l) \ll \tau_{\alpha S}(k_p; T_l)$ and equilibrate to slightly lower temperatures before $b(t) \to 0$ halts $u(t)$ and freezes all modes simultaneously.

Finally, Fig.~\ref{fig:fictive_temperature}(c) shows the mutual consistency of the limiting fictive temperatures $T'_f(q)$ across $q \in [10^{-4}, 10^2]$ for reciprocal-space ($T'^{[S_{\max}]}_f$, $T_f^{\text{TNM}}(k_p)$), real-space ($T'^{[g_2]}_f$, $T'^{[g_1]}_f$), and entropic ($T'^{[s_2]}_f$) observables, all bending concertedly toward $T_c = 0.2287$ as $q \to 0$. 
The cage-scale estimators $T'^{[S_{\max}]}_f$, $T'^{[g_2]}_f$, and $T_f^{\text{TNM}}(k_p)$ nearly coincide across all seven decades, whereas the slight downward offset of short-range observables reflects the faster equilibration of high-$k$ modes in Fig.~\ref{fig:fictive_temperature}(b)~\cite{XinLi2017, McKenna2020, DiLisio2023}. 
Given this global consistency, we henceforth adopt $T_f(T(t); q) \equiv T_f^{[S_{\max}]}(T(t); q)$ as our canonical fictive temperature.

\section{Non-Equilibrium Kauzmann Diagrams and Relaxational Heat Capacity upon Cooling}\label{sec:KauzmannDiagrams}

\subsection{Microscopic Two-Body Excess Entropy and Dynamic Resolution of the Kauzmann Paradox}\label{subsec:TwoBodyEntropy}

Having recovered both the instantaneous spatial correlations $g(r; u(t); q)$ and the canonical structural fictive temperature $T_f(T(t); q) \equiv T_f^{[S_{\max}]}(T(t); q)$ along continuous cooling ramps, we are now in a position to construct non-equilibrium Kauzmann diagrams directly from microscopic liquid-state theory. 
In its classical formulation~\cite{Kauzmann1948, Angel1995, Debenedetti2001, Berthier2009, Mauro2009}, the Kauzmann paradox envisions an entropy crisis when the configurational entropy of a supercooled liquid $S_c(T) = S_{\text{liq}}(T) - S_{\text{cryst}}(T)$ is extrapolated below the kinetic glass transition, vanishing at a finite temperature $T_K > 0$. 
For a model fluid governed by soft-core repulsive interactions under the constraint of macroscopic homogeneity, crystal nucleation pathways are suppressed, and the natural configurational ground state of the amorphous branch is geometrically bounded by the random close packing (RCP) limit $\phi_{\text{RCP}} \approx 0.64$~\cite{HansenMcDonald1990, Truskett2000, Berthier2009}. 
Within the pair-correlation framework of liquid-state theory~\cite{HansenMcDonald1990, Pond2011}, the structural loss of accessible configurations is convincingly mirrored by the two-body excess entropy per particle $s_2(u(t); q)/k_B$, which captures more than $85\%$ of the total excess entropy in dense simple liquids. 
Accordingly, the temperature $T_K(\phi)$ at which the equilibrium effective packing fraction attains random close packing,
\begin{equation}\label{eq:TK_definition}
\phi_{\text{eff}}^{\text{eq}}(\phi, T_K) = \phi\,\lambda^3(\phi, T_K) = \phi_{\text{RCP}} = 0.64,
\end{equation}
defines an operational packing-crisis Kauzmann point for the metastable disordered fluid, below which the system would lose all configurational phase space.
At $\phi = 0.70$ ($\nu = 6$), Eq.~\eqref{eq:TK_definition} gives $T_K = 0.0525$ ($T_K/T_c = 0.230$), where the equilibrium two-body excess entropy reaches $s_2^{\text{eq}}(T_K)/k_B = -10.892$ compared to $s_2^{\text{eq}}(T_c)/k_B = -7.308$ at the dynamic arrest transition $T_c = 0.2287$.

\begin{figure}[t!]
    \centering
    \includegraphics[width=0.45\textwidth]{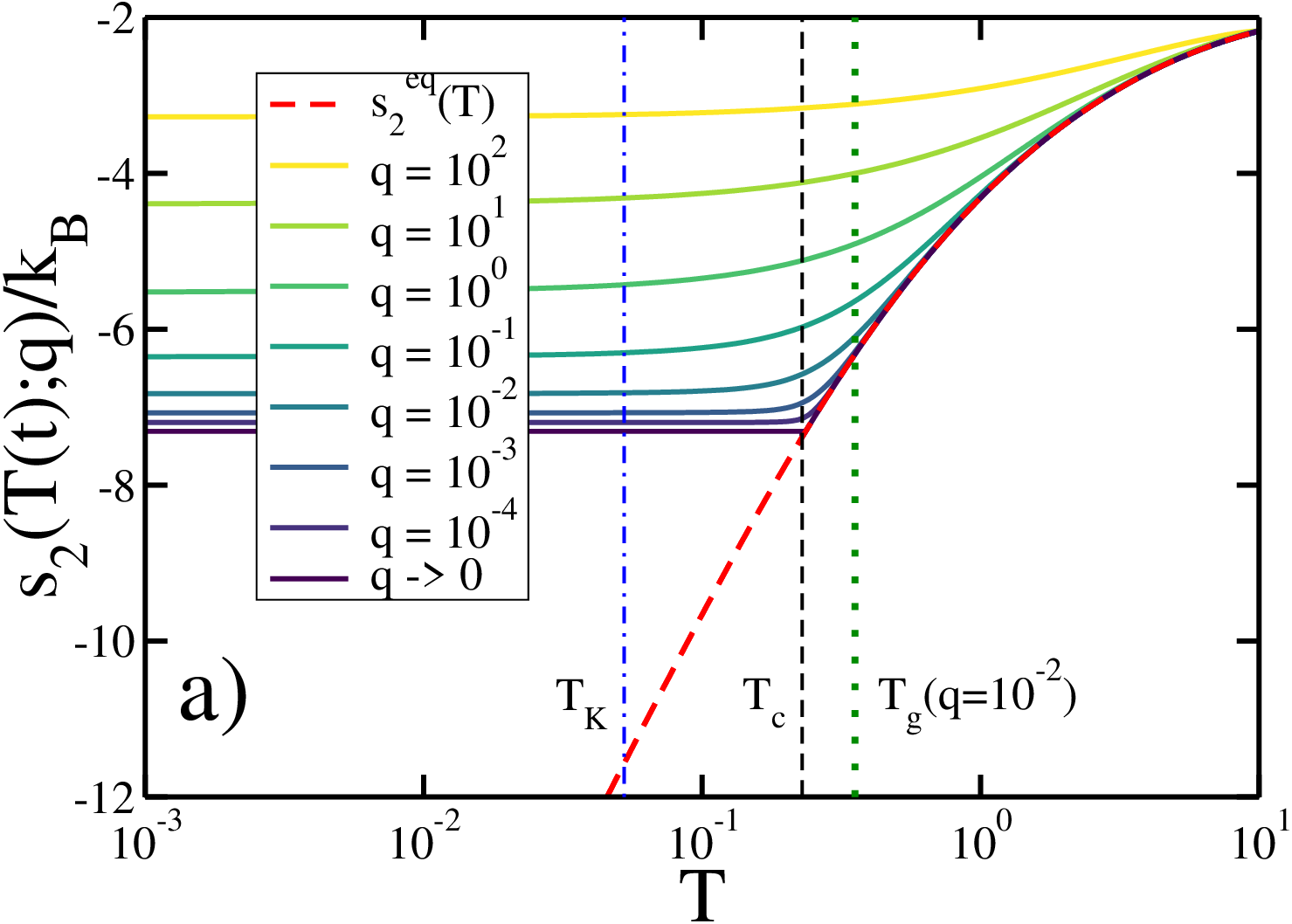}
    \includegraphics[width=0.45\textwidth]{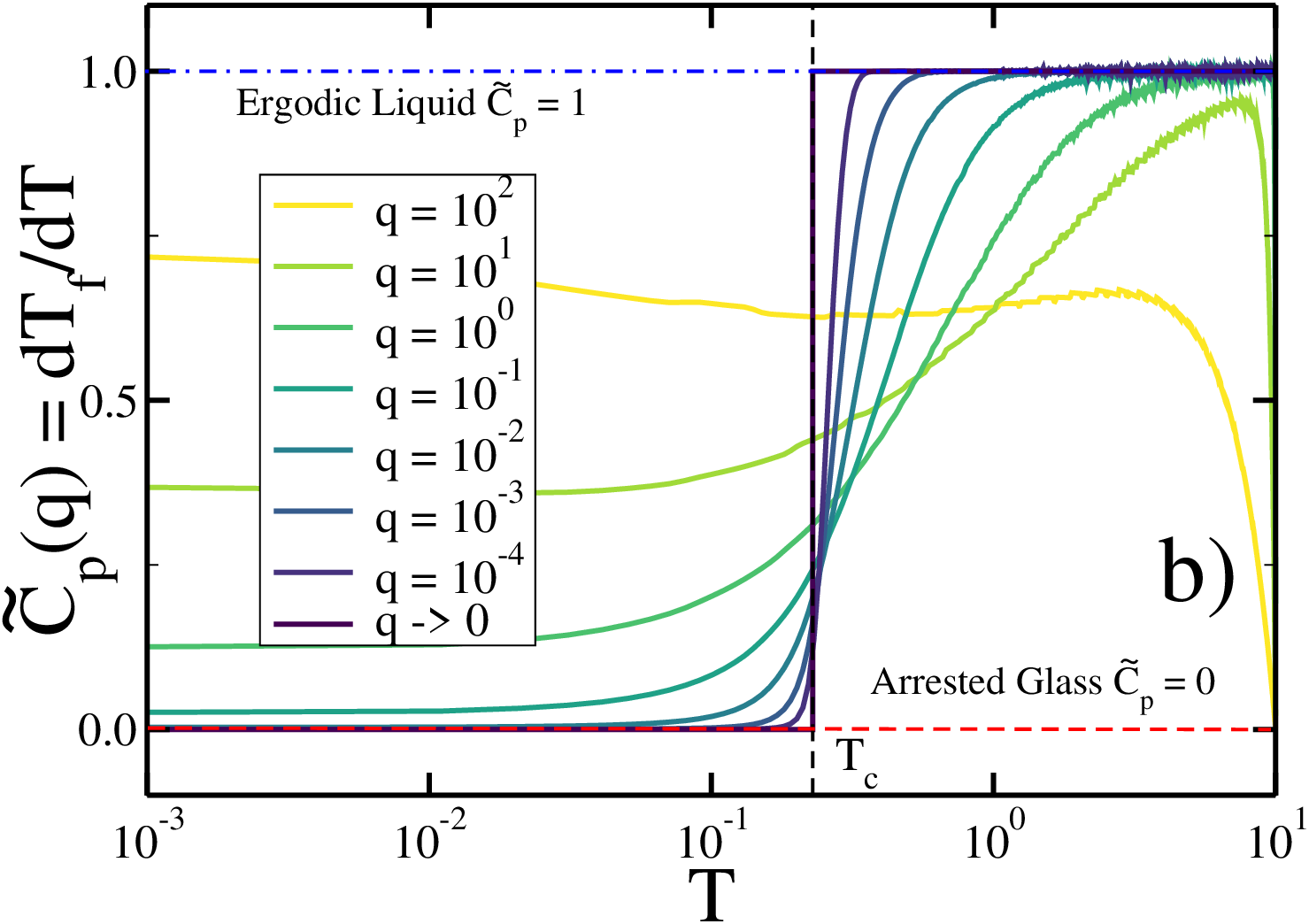}
    \caption{
    \textbf{(a)} Instantaneous two-body excess entropy $s_2(T(t); q)/k_B$ computed from $g(r; u(t); q)$ via Eq.~\eqref{eq:s2_definition} vs instantaneous bath temperature $T(t) = T_i - q t$, compared against the Ornstein--Zernike equilibrium branch $s_2^{\text{eq}}(T)/k_B$ (red dashed curve) and the analytical quasi-static limit $q \to 0$ [Eq.~\eqref{eq:s2_quasistatic_limit}, dark purple solid floor at $s_2^{\text{eq}}(T_c)/k_B = -7.308$], with vertical lines marking the equilibrium Kauzmann point $T_K = 0.0525$ (blue dash-dotted), the critical arrest temperature $T_c = 0.2287$ (black dashed), and the representative glass transition temperature $T_g(q = 10^{-2}) = 0.354$ (green dotted). \textbf{(b)} Cooling-ramp normalized relaxational heat capacity $\tilde{C}_p(T(t); q) = \mathrm{d}T_f(T(t); q)/\mathrm{d}T(t)$ vs $T(t) = T_i - q t$, illustrating the rate-dependent sigmoidal vitrification step from the ergodic liquid plateau ($\tilde{C}_p = 1$, blue dash-dotted line) down to the arrested glass baseline ($\tilde{C}_p = 0$, red dashed line) and its convergence toward the analytical quasi-static Heaviside step $\Theta(T - T_c)$ [Eq.~\eqref{eq:Cp_quasistatic_limit}, dark purple solid step] at $T_c = 0.2287$ (vertical black dashed line).}
    \label{fig:kauzmann_diagrams}
\end{figure}

Figure~\ref{fig:kauzmann_diagrams}(a) compares the evolution of $s_2(T(t); q)/k_B$ along the seven NESCGLE cooling ramps against the equilibrium OZE curve $s_2^{\text{eq}}(T)/k_B$ as a function of the instantaneous bath temperature $T(t) = T_i - q t$. 
Along the equilibrium branch (red dashed curve), the two-body excess entropy decreases monotonically upon cooling, crossing the dynamic arrest temperature $T_c = 0.2287$ (vertical black dashed line) and plunging directly toward the Kauzmann catastrophe at $T_K = 0.0525$ (vertical blue dash-dotted line). 
In stark contrast, along every finite-rate cooling ramp ($q > 0$), the progressive deceleration and finite arrest of the microscopic inner clock ($u(t) \to u^a(q)$) causes the non-equilibrium liquid to peel away from the equilibrium curve near its kinetic glass transition temperature [illustrated by the vertical green dashed line at $T_g(q = 10^{-2}) = 0.354$, see Appendix \ref{app:TgConstructions}] and freeze onto a horizontal residual-entropy plateau. 
Under an ultra-fast quench ($q = 10^2$), the system arrests in a weakly correlated fluid configuration with $s_2(t_{\text{end}})/k_B = -3.275$. 
As the cooling rate is slowed across seven orders of magnitude down to $q = 10^{-4}$, this frozen plateau descends systematically to $-7.196$, confining all vitrified states to strictly positive residual entropies trapped just above the ideal arrest threshold $T_c$. 

As established in Sec.~\ref{subsec:Sk_and_gr}, the exact lower bound of this cooling-rate progression is governed by the analytical quasi-static limit $q \to 0^+$, in which the pair structure tracks the Ornstein--Zernike equilibrium branch down to $T_c$ and arrests permanently in the critical configuration $g^{\text{eq}}(r; T_c)$ for all $T(t) \le T_c$ once $b_{\text{eq}}(T \le T_c) = 0$. 
Consequently, as $q \to 0^+$, the non-equilibrium two-body excess entropy converges pointwise to the piecewise-continuous lower envelope [dark purple solid curve in Fig.~\ref{fig:kauzmann_diagrams}(a)]:
\begin{equation}\label{eq:s2_quasistatic_limit}
\lim_{q \to 0^+} \frac{s_2(T(t); q)}{k_B} = \begin{cases} s_2^{\text{eq}}(T)/k_B, & T(t) \ge T_c, \\[4pt] s_2^{\text{eq}}(T_c)/k_B = -7.308, & T(t) < T_c, \end{cases}
\end{equation}
which exhibits a sharp slope discontinuity at $T_c = 0.2287$ and remains strictly bounded above the Kauzmann catastrophe at $T_K = 0.0525$ ($s_2^{\text{eq}}(T_K)/k_B = -10.892$)~\cite{Kauzmann1948, Mauro2009, Cangialosi2023}. 
It should be noted that within the ideal SCGLE and NESCGLE frameworks, $T_c$ represents a critical mean-field dynamic arrest boundary analogous to the ideal mode-coupling transition temperature $T_{\text{MCT}}$~\cite{Goetze2009, Gory2019}. 
In laboratory experiments on molecular and colloidal glass-formers, thermally activated hopping processes eventually bypass this ideal singularity and restore ergodicity below $T_c$, shifting the measured laboratory glass transition temperature $T_g^{\text{exp}}$ below the ideal arrest threshold ($T_c > T_g^{\text{exp}}$)~\cite{Berthier2009, Gory_prl2022}. 
In the absence of hopping, however, $T_c$ acts as an impassable kinetic boundary: 
upon cooling from the ergodic liquid, the progressive deceleration of the inner clock causes the system to fall out of equilibrium at a kinetic vitrification temperature $T_g(q)$ that lies strictly above $T_c$ ($T_g(q) > T_c$). Consequently, the Kauzmann catastrophe at $T_K < T_c$ remains dynamically inaccessible for any finite cooling rate $q > 0$.

\subsection{Cooling-Ramp Normalized Relaxational Heat Capacity}\label{subsec:CoolingHeatCapacity}

Beyond the integral excess entropy $s_2$, the continuous cooling ramps provide direct access to the non-equilibrium calorimetric response of the vitrifying liquid~\cite{Moynihan1976, Yue2004, Weyer2005, Cangialosi2023}. 
Within the Tool--Narayanaswamy--Moynihan (TNM) framework~\cite{Tool1946, Narayanaswamy1971, Moynihan1976}, the macroscopic thermal response of an out-of-equilibrium glass-forming liquid is governed by the fundamental separation of time scales between fast intra-cage particle vibrations and slow collective configurational rearrangements. 
In standard Differential Scanning Calorimetry (DSC) nomenclature~\cite{Moynihan1976} (where the subscript $p$ denotes isobaric laboratory conditions), $C_p(T)$ represents the instantaneous non-equilibrium heat capacity measured along the thermal ramp at bath temperature $T(t)$, $C_{pg}(T)$ (subscript $g$ for \emph{glass}) denotes the vibrational heat-capacity baseline of the arrested glass in which configurational rearrangements are completely frozen, and $C_{pe}(T)$ (subscript $e$ for \emph{equilibrium liquid}) denotes the heat capacity of the fully equilibrated liquid, which includes both the fast intra-cage vibrational contribution $C_{pg}(T)$ and the configurational contribution $\Delta C_p(T) \equiv C_{pe}(T) - C_{pg}(T)$ arising from structural rearrangements. 
While short-time intra-cage vibrations equilibrate quasi-instantaneously with the external thermal bath at $T(t)$ to establish $C_{pg}(T)$, slow cooperative cage-breaking rearrangements evolve solely on the timescale of the microscopic inner clock $u(t)$ and are parameterized by Tool's structural fictive temperature $T_f(T(t); q)$. 
Subtracting the vibrational glass baseline $C_{pg}(T)$ and normalizing by the equilibrium configurational step $C_{pe}(T) - C_{pg}(T)$ therefore isolates the dimensionless active fraction of the configurational heat capacity, which Moynihan~\cite{Moynihan1976, Weyer2005, Gujrati2009} termed the normalized relaxational specific heat $\tilde{C}_p(T(t); q)$ and identified directly with the temperature derivative of the structural fictive temperature (retaining the conventional symbol $\tilde{C}_p$ here at fixed volume fraction $\phi$ for direct comparison with TNM calorimetry):
\begin{equation}\label{eq:Cp_tilde_def}
\tilde{C}_p(T(t); q) \equiv \frac{C_p(T) - C_{pg}(T)}{C_{pe}(T) - C_{pg}(T)} = \left(\frac{\partial T_f(T(t); q)}{\partial T(t)}\right)_{\phi,q} = \frac{\mathrm{d}T_f(T(t); q)}{\mathrm{d}T(t)}.
\end{equation}
In the high-temperature ergodic liquid regime where structural relaxation occurs instantaneously on the laboratory timescale ($T_f(T(t); q) = T(t)$ and $C_p(T) = C_{pe}(T)$), Eq.~\eqref{eq:Cp_tilde_def} yields $\tilde{C}_p(T(t); q) = 1$, whereas deep in the vitrified state where the structure is frozen ($T_f(T(t); q) = T'_f(q)$ and $C_p(T) = C_{pg}(T)$), the relaxational specific heat vanishes ($\tilde{C}_p(T(t); q) = 0$). 
Alongside $\tilde{C}_p(T(t); q)$, our microscopic formulation also defines the unnormalized two-body pair-correlation excess heat capacity per particle $C_2^{\text{ex}}(T(t); q)/k_B$ directly from the evolving spatial correlations by differentiating the microscopic two-body excess entropy $s_2(T(t); q)/k_B$ of Eq.~\eqref{eq:s2_definition} with respect to the bath temperature:
\begin{equation}\label{eq:C2_and_Cpconf_def}
\frac{C_2^{\text{ex}}(T(t); q)}{k_B} \equiv T(t)\,\frac{\mathrm{d}\left[s_2(T(t); q)/k_B\right]}{\mathrm{d}T(t)}.
\end{equation}
It is worth emphasizing that $C_2^{\text{ex}}(T(t); q)/k_B$ in Eq.~\eqref{eq:C2_and_Cpconf_def} isolates exclusively the structural (configurational) contribution associated with interparticle pair correlations. 
In experimental Differential Scanning Calorimetry (DSC) thermograms, this macroscopic response corresponds directly to the active configurational heat-capacity increment, $\Delta C_p(T) \equiv C_p(T) - C_{pg}(T)$. 
Here, the glassy baseline $C_{pg}(T) > 0$ --which microscopically originates from fast intra-cage vibrational phonon modes-- has been formally subtracted to isolate the slow structural rearrangements. 
Consequently, as the microscopic inner clock arrests ($b(t) \to 0$), structural rearrangements freeze out, causing $C_2^{\text{ex}}$ to vanish at the lowest temperatures in exact correspondence with the normalized relaxational specific heat baseline $\tilde{C}_p(T(t); q) \to 0$.

Figure~\ref{fig:kauzmann_diagrams}(b) shows $\tilde{C}_p(T(t); q)$ across $q \in [10^{-4}, 10^2]$ and $q \to 0$. 
For $q \in [10^{-4}, 10^0]$, the liquid starts on the ergodic plateau $\tilde{C}_p = 1$ (blue dash-dotted) and undergoes a sigmoidal calorimetric glass transition step down to the glass baseline $\tilde{C}_p = 0$ (red dashed) as the inner clock decelerates above $T_c = 0.2287$ (vertical black dashed line). Decreasing $q$ from $10^0$ to $10^{-4}$ shifts this vitrification step toward $T_c$ and narrows its width. 
Since $\lim_{q \to 0^+} T_f(T(t); q) = \max(T(t), T_c)$, the quasi-static limit of $\tilde{C}_p(T(t); q)$ is an exact Heaviside step at $T_c$ [dark purple solid curve in Fig.~\ref{fig:kauzmann_diagrams}(b)]:
\begin{equation}\label{eq:Cp_quasistatic_limit}
\lim_{q \to 0^+} \tilde{C}_p(T(t); q) = \Theta(T(t) - T_c) = \begin{cases} 1, & T(t) > T_c, \\[4pt] 0, & T(t) < T_c. \end{cases}
\end{equation}
Equation~\eqref{eq:Cp_quasistatic_limit} shows that in the quasi-static limit $q \to 0^+$, the continuous calorimetric vitrification crossover sharpens into an ideal discontinuous Ehrenfest-like heat-capacity step at $T_c = 0.2287$~\cite{DaviesJones1953, Gujrati2009}. 
Conversely, under the fastest cooling rates ($q = 10^1$ and $10^2$), the elementary residence time $\Delta t = \Delta T/q$ is already shorter than or comparable to the structural relaxation time near $T_i = 10.0$, so $\tilde{C}_p(T(t); q)$ departs from unity immediately upon cooling, passing through a broad intermediate plateau ($\tilde{C}_p \approx 0.65\text{--}0.70$ for $q = 10^2$) before dropping toward zero. 
Having characterized the non-equilibrium Kauzmann diagrams and the relaxational heat capacity upon continuous cooling, we now synthesize the operational extraction of the calorimetric glass transition temperature $T_g(q)$ and examine its critical scaling behavior as $q \to 0$.

\section{Calorimetric Glass Transition Temperature and Critical Cooling-Rate Scaling}\label{sec:CalorimetricTg}

\subsection{Operational Integral and Differential Calorimetric Definitions of $T_g(q)$ upon Cooling}\label{subsec:TgOperational}

In experimental differential scanning calorimetry and dilatometry, the glass transition temperature $T_g(q)$ is operationally extracted from continuous cooling thermograms through two complementary approaches. 
The first relies on integral equal-area constructions that identify an ideal thermodynamic step conserving enthalpy or configurational entropy across the vitrification interval~\cite{Moynihan1976, Hodge1994, Yue2004, Weyer2005, Gujrati2009}. 
Complementarily, differential criteria identify characteristic landmarks directly on the heat-capacity trace, such as the vitrification onset peak or the half-step midpoint~\cite{Bruning1992, Angel1995, Vollmayr1996, Debenedetti2001, Cangialosi2023}.
To demonstrate the internal consistency of our first-principles framework across both experimental conventions, we evaluate four operational cooling-ramp definitions of $T_g(q)$ over the entire kinetic spectrum $q \in [10^{-4}, 10^2]$ (explicit step-by-step graphical constructions for each of the four criteria at a representative cooling rate $q = 10^{-2}$ are detailed in Appendix~\ref{app:TgConstructions} and illustrated in Fig.~\ref{fig:app_Tg_constructions}).

The two integral equal-area definitions of $T_g(q)$ upon cooling are given by
\begin{equation}\label{eq:Tg_calorimetric_integrals}
T_g^{\text{Moy}}(q) \equiv T_i - \int_{T_{\text{end}}}^{T_i} \tilde{C}_p(T; q)\,\mathrm{d}T = T'_f(q), \qquad T_g^{[C_2]}(q) \equiv T_i - \int_{T_{\text{end}}}^{T_i} \frac{C_2^{\text{ex}}(T; q)}{C_2^{\text{ex,eq}}(T)}\,\mathrm{d}T.
\end{equation}
Here, $T_g^{\text{Moy}}(q)$ represents Moynihan's canonical equal-area construction applied to the normalized relaxational heat capacity $\tilde{C}_p(T; q) = \mathrm{d}T_f/\mathrm{d}T$. 
Along the cooling ramp, this construction balances the deficit area $A^+$ ($\tilde{C}_p < 1$ for $T > T_g^{\text{Moy}}$) against the residual tail area $A^-$ ($\tilde{C}_p > 0$ for $T < T_g^{\text{Moy}}$) [Fig.~\ref{fig:app_Tg_constructions}(a)]. 
By the fundamental theorem of calculus, the resulting transition temperature coincides identically with the limiting glass fictive temperature, $T_g^{\text{Moy}}(q) = T'_f(q) \equiv T_f(T_{\text{end}}; q)$, which geometrically corresponds to the intersection between the extrapolated glass plateau $T_f = T'_f(q)$ and the equilibrium liquid line $T_f = T$.
Complementarily, $T_g^{[C_2]}(q)$ in Eq.~\eqref{eq:Tg_calorimetric_integrals} applies the equal-area step construction directly to the microscopic two-body pair-correlation heat-capacity ratio $C_2^{\text{ex}}(T; q)/C_2^{\text{ex,eq}}(T)$ upon cooling, where $A^+ = A^-$, providing an entropic integral definition independent of the TNM fictive-temperature mapping. 
Alongside these two integral measures, we also extract two direct differential landmarks from the unnormalized cooling-ramp excess heat capacity profile $C_2^{\text{ex}}(T(t); q)/k_B$ [Fig.~\ref{fig:app_Tg_constructions}(b)]: 
(i) the peak temperature $T_{\max}^{[C_2]}(q)$ at which $C_2^{\text{ex}}(T(t); q)/k_B$ attains its global maximum $C_{2,\max}^{\text{ex}}(q)/k_B$ as the cooling trajectory peels away from the rising Ornstein--Zernike equilibrium curve $C_2^{\text{ex,eq}}(T)/k_B$ (marking the high-temperature onset of the calorimetric glass transition step), and (ii) the half-step midpoint temperature $T_{g,1/2}^{[C_2]}(q) < T_{\max}^{[C_2]}(q)$ on the low-temperature vitrification flank where $C_2^{\text{ex}}(T_{g,1/2}^{[C_2]}; q) = \frac{1}{2}\,C_{2,\max}^{\text{ex}}(q)$.

\begin{figure}[t!]
    \centering
    \includegraphics[width=0.5\textwidth]{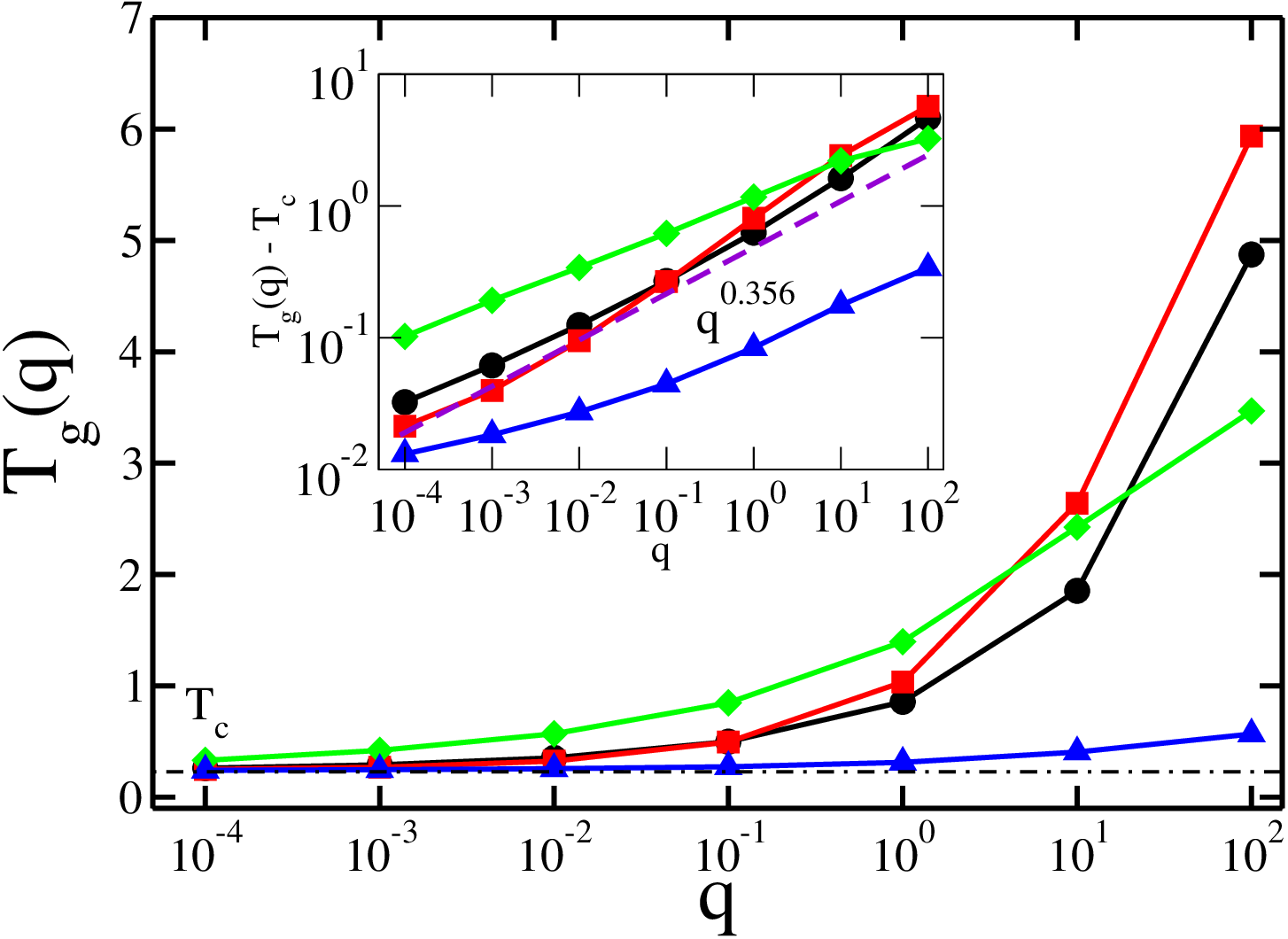}
    \caption{
    Cooling-rate dependence of the four operational calorimetric definitions of the glass transition temperature---Moynihan's equal-area temperature $T_g^{\text{Moy}}(q) = T'_f(q)$ [Eq.~\eqref{eq:Tg_calorimetric_integrals}, black filled circles], the microscopic two-body equal-area temperature $T_g^{[C_2]}(q)$ [Eq.~\eqref{eq:Tg_calorimetric_integrals}, red filled squares], the cooling heat-capacity peak onset temperature $T_{\max}^{[C_2]}(q)$ (green filled diamonds), and the half-step vitrification midpoint $T_{g,1/2}^{[C_2]}(q)$ (blue filled triangles)---relative to the horizontal asymptotes at the ideal arrest temperature $T_c = 0.2287$. Inset: Double-logarithmic plot of the excess glass transition temperature $T_g(q) - T_c$ versus cooling rate $q$, demonstrating the critical power-law scaling $T_g(q) - T_c \propto q^{\psi}$ ($\psi \approx 0.356$, purple dashed line) across all four calorimetric definitions.}
    \label{fig:calorimetric_Tg}
\end{figure}

Figure~\ref{fig:calorimetric_Tg} displays the resulting cooling-rate dependence of all four calorimetric glass transition temperatures across the entire kinetic spectrum $q \in [10^{-4}, 10^2]$. 
Regardless of whether $T_g(q)$ is defined via the canonical Moynihan integral $T_g^{\text{Moy}}(q)$ (black filled circles), the two-body heat-capacity equal-area integral $T_g^{[C_2]}(q)$ (red filled squares), the calorimetric onset peak $T_{\max}^{[C_2]}(q)$ (green filled diamonds), or the half-step vitrification midpoint $T_{g,1/2}^{[C_2]}(q)$ (blue filled triangles), all four determinations decrease strictly monotonically with decreasing $q$ and remain bounded from below by the ideal dynamic arrest temperature $T_c = 0.2287$ (horizontal black dashed line), well above the Kauzmann catastrophe at $T_K = 0.0525$. 
Moreover, a clear and physically transparent hierarchy emerges across the supercooled regime: the two integral equal-area definitions $T_g^{\text{Moy}}(q)$ and $T_g^{[C_2]}(q)$ track each other closely over all seven orders of magnitude in cooling rate and are consistently bracketed from above by the high-temperature onset peak $T_{\max}^{[C_2]}(q)$ (where the cooling trajectory first peels away from the equilibrium liquid curve) and from below by the half-step midpoint $T_{g,1/2}^{[C_2]}(q)$ (where structural arrest is already half-complete). 
As $q$ decreases toward the quasi-static limit, the width of the glass transformation interval narrows and all four operational DSC conventions converge toward the same underlying asymptotic arrest boundary $T_c$, confirming that each protocol probes a distinct stage of the same inner-clock vitrification crossover.

The lower bound $T_g(q) \ge T_c$ obeyed by all four calorimetric definitions in Fig.~\ref{fig:calorimetric_Tg} reflects the fundamental kinematics of continuous cooling across an ideal mean-field arrest boundary. 
Because the structural relaxation time diverges algebraically at $T_c$ in the absence of activated hopping, any finite cooling rate $q > 0$ inevitably triggers kinetic vitrification at an earlier stage of the ramp where structural relaxation first falls out of pace with the thermal bath, enforcing the theoretical hierarchy $T_g(q) > T_c$. 
Only in the quasi-static cooling limit ($q \to 0^+$), where the system is granted an infinite chronological waiting time to exhaust structural equilibration down to the singularity, does the kinetic vitrification temperature converge identically to the dynamic arrest boundary, $\lim_{q \to 0^+} T_g(q) = T_c$. 
This behavior highlights that while activated processes in real glass-formers typically yield $T_c > T_g^{\text{exp}}$, the idealized mean-field framework treats $T_c$ as the absolute lower temperature horizon for continuous liquid vitrification.

\subsection{Critical Power-Law Scaling and Apparent Moynihan Activation Enthalpy}\label{subsec:TgCriticalScaling}

To understand the physical origin of the pronounced upward curvature of $T_g(q)$ versus $\log_{10}q$ in the main panel of Fig.~\ref{fig:calorimetric_Tg}, we derive the asymptotic scaling of the glass transition temperature directly from the NESCGLE inner-clock differential equation $\mathrm{d}u = b(t)\,\mathrm{d}t$. 
Near the critical dynamic arrest boundary ($T \to T_c^+$), the equilibrium SCGLE collective mobility vanishes algebraically according to the mode-coupling power law $b_{\text{eq}}(T) \propto \tau_\alpha^{-1}(T) \propto (T - T_c)^{\gamma}$ with critical exponent $\gamma \approx 1.90$~\cite{Goetze2009, Juarez2007, Gory2019}. 
Along a constant-rate cooling ramp $\mathrm{d}T = -q\,\mathrm{d}t$, kinetic vitrification sets in when the remaining material-time increment that can be accumulated before reaching the dynamic arrest boundary $T_c$ becomes of order unity, $\Delta u_{\text{rem}} = \int_{T_c}^{T_g(q)} [b_{\text{eq}}(T)/q]\,\mathrm{d}T \sim \mathcal{O}(1)$. 
Integrating the asymptotic mobility $b_{\text{eq}}(T) \propto (T - T_c)^{\gamma}$ directly yields the critical power-law scaling relation
\begin{equation}\label{eq:Tg_power_law_scaling}
T_g(q) - T_c = A\,q^{\psi}, \qquad \psi = \frac{1}{\gamma + 1}.
\end{equation}

The double-logarithmic inset of Fig.~\ref{fig:calorimetric_Tg} tests Eq.~\eqref{eq:Tg_power_law_scaling} by plotting the excess glass transition temperature $T_g(q) - T_c$ versus $q$. 
Across all seven orders of magnitude in cooling rate, the calorimetric indicators straighten into power-law trajectories in the double-logarithmic plane. 
Specifically, Moynihan's canonical equal-area temperature follows Eq.~\eqref{eq:Tg_power_law_scaling} with an exponent $\psi \approx 0.358$ across all seven decades (illustrated by the purple dashed line $\propto q^\psi$ in the inset, and $\psi \approx 0.339$ across the supercooled regime $q \le 10^1$), in quantitative agreement with the analytical inner-clock scaling prediction $\psi = 1/(\gamma + 1) \approx 0.345$ [Eq.~\eqref{eq:Tg_power_law_scaling}]. 
Crucially, differentiating Eq.~\eqref{eq:Tg_power_law_scaling} yields the apparent Moynihan activation enthalpy extracted in experimental cooling-rate shift-factor analyses~\cite{Moynihan1976, Bruning1992, Hodge1994, Vollmayr1996, Yue2004, Cangialosi2023},
\begin{equation}\label{eq:Moynihan_apparent_enthalpy}
\frac{\Delta H_{\text{app}}^*(T_g)}{k_B} \equiv -\frac{\mathrm{d}\ln q}{\mathrm{d}(1/T_g)} = \frac{T_g^2}{\psi\,(T_g - T_c)},
\end{equation}
which increases rapidly as $q$ is reduced and $T_g(q)$ approaches $T_c$ from above. 
Equations~\eqref{eq:Tg_power_law_scaling} and~\eqref{eq:Moynihan_apparent_enthalpy} thus establish a direct analytical bridge between the fragile super-Arrhenius growth of the apparent calorimetric activation barrier $\Delta H_{\text{app}}^*(T_g)$ and the critical power-law divergence of the structural relaxation time at $T_c$.

\section{Conclusions and Perspectives}\label{sec:Conclusions}

In this work, we have established a first-principles statistical-mechanical bridge between the microscopic theory of non-equilibrium structural relaxation and the macroscopic Tool--Narayanaswamy--Moynihan (TNM) phenomenology of glass-forming liquids under continuous cooling.
Rooted in the canonical relaxation equations of the non-equilibrium Onsager--Machlup theory of thermal fluctuations~\cite{Onsager1931_1, Onsager1931_2, Onsager1953, Machlup1953, Keizer1987, Pedro2010, Peredo2024, MedinaNoyola2026AnnRev}, the Non-Equilibrium Self-Consistent Generalized Langevin Equation (NESCGLE) theory provides a microscopic interpretation of Narayanaswamy's internal material clock~\cite{Narayanaswamy1971, Peredo2022}: 
the accumulated collective particle mobility $u(t) = \int_0^t b(t')\,\mathrm{d}t'$ linearizes the time evolution of every spatial Fourier mode of the non-equilibrium static structure factor $S(k; u(t))$. 
By coupling this microscopic inner-clock formulation with the piece-wise continuous thermal-ramp formalism~\cite{Johnathan2019}, we have investigated continuous cooling across seven decades of thermal rate ($q \in [10^{-4}, 10^2]$, with $M = 1000$ steps) alongside the analytical quasi-static limit ($q \to 0$) for a benchmark Weeks--Chandler--Andersen (WCA) soft-sphere liquid ($\nu = 6$, $\phi = 0.70$). 
This theoretical framework enables us to recover Tool's fictive temperature $T_f(T(t); q)$ directly from evolving spatial pair correlations, construct self-consistent non-equilibrium Kauzmann diagrams and cooling-ramp relaxational heat capacities, and establish the operational extraction and critical cooling-rate scaling of the glass transition temperature $T_g(q)$.
Tracking the instantaneous static structure factor $S(k; u(t); q)$ and radial distribution function $g(r; u(t); q)$ parameterized by the laboratory bath temperature $T(t)$ and benchmarking them against the exact Ornstein--Zernike equilibrium reference branch demonstrated that a liquid cooled at any finite rate $q > 0$ never reaches the equilibrium target ($T_{\text{end}} = 10^{-3}$). 
Instead, as the inner clock decelerates ($\mathrm{d}u = b(t)\,\mathrm{d}t \to 0$) upon approaching the critical dynamic arrest temperature $T_c = 0.2287$, the microstructure freezes into a non-equilibrium glassy state confined within a strictly bounded structural arrest corridor between the high-temperature fluid configuration ($q = 10^2$) and the ideal arrest boundary ($q = 10^{-4}$ and the quasi-static limit $q \to 0$, which arrests sharply at $T_c = 0.2287$). 
Calibrating structural features against their equilibrium temperature calibration curves via horizontal projection recovers Tool's fictive temperature $T_f(T(t); q)$. 
Furthermore, the explicit wavenumber dependence of the NESCGLE relaxation rate $\tau_{\alpha S}^{-1}(k; t) = b(t)\,\Gamma(k; T)$ reveals that an arrested glass is characterized not by a single scalar fictive temperature, but by a continuous mode fictive-temperature spectrum $T_f(k; u(t); q)$ that peaks sharply at the primary cage wavenumber $k_p\sigma \approx 7.12$ due to de Gennes narrowing~\cite{deGennes1959} ($\mathcal{E}(k_p; T) = 1/S^{\text{eq}}(k_p; T) \ll 1$). 
This multi-mode dispersion naturally explains the experimental observation that distinct macroscopic observables measured along the exact same thermal history freeze at slightly different fictive temperatures~\cite{XinLi2017, McKenna2020, Cangialosi2023, DiLisio2023}, while motivating the primary diffraction-peak fictive temperature $T_f(T(t); q) \equiv T_f^{[S_{\max}]}(T(t); q)$ as the canonical internal state parameter of the vitrified cage structure.

Evaluating the two-body excess entropy $s_2(u(t); q)/k_B$ and the normalized relaxational specific heat $\tilde{C}_p(T(t); q) = \mathrm{d}T_f(T(t); q)/\mathrm{d}T(t)$ along continuous cooling ramps enabled a first-principles construction of non-equilibrium Kauzmann diagrams and a transparent dynamical resolution of the Kauzmann paradox~\cite{Kauzmann1948, Debenedetti2001, Mauro2009, McKenna2020}. 
Whereas extrapolating the equilibrium Ornstein--Zernike liquid branch below $T_c = 0.2287$ leads to a geometric random-close-packing catastrophe at a finite Kauzmann temperature $T_K = 0.0525$ ($T_K/T_c = 0.23$, where $\phi_{\text{eff}}^{\text{eq}}(T_K) = \phi_{\text{RCP}} = 0.64$), the vanishing of the collective mobility $b(t) \to 0$ arrests every finite cooling ramp above $T_c > T_K$, trapping the system on a residual-entropy plateau and rendering $T_K$ kinetically inaccessible across all seven decades of cooling rate. 
Simultaneously, the cooling-ramp relaxational heat capacity $\tilde{C}_p(T(t); q)$ captures the rate-dependent sigmoidal vitrification step from the ergodic liquid plateau ($\tilde{C}_p = 1$) down to the arrested glass baseline ($\tilde{C}_p = 0$), sharpening toward an ideal Heaviside step $\Theta(T - T_c)$ in the quasi-static limit $q \to 0^+$.

Finally, synthesizing the calorimetric response upon continuous cooling allowed us to extract the glass transition temperature $T_g(q)$ across four operational criteria: Moynihan's canonical equal-area construction $T_g^{\text{Moy}}(q) = T'_f(q)$, the microscopic two-body heat-capacity equal-area integral $T_g^{[C_2]}(q)$, the cooling heat-capacity onset peak $T_{\max}^{[C_2]}(q)$, and the half-step vitrification midpoint $T_{g,1/2}^{[C_2]}(q)$.
Across all four definitions, $T_g(q)$ decreases strictly monotonically over seven orders of magnitude in cooling rate and approaches the ideal dynamic arrest boundary according to the critical power law $T_g(q) - T_c = A\,q^{\psi}$ with $psi = 1/(\gamma + 1) \approx 0.358$. 
This scaling relation provides a direct microscopic rationale for both the super-Arrhenius curvature of $T_g(q)$ and the diverging apparent activation enthalpy $\Delta H_{\text{app}}^*(T_g) \propto (T_g - T_c)^{-1}$ observed in calorimetric cooling-rate studies, tracing them back to the critical deceleration of collective particle mobility as the dynamic arrest boundary is approached.

The theoretical framework established in this work lays the foundation for several immediate developments. 
First and foremost, having characterized the primary liquid-to-glass vitrification crossover under monotonic cooling, standard Differential Scanning Calorimetry (DSC) protocols routinely pair continuous cooling with subsequent reheating across closed thermal cycles ($T_i \to T_{\text{min}} \to T_i$ at controlled rates $\mp q$). In an upcoming paper, we extend the present inner-clock and fictive-temperature methodology to closed thermal hysteresis cycles, demonstrating how the NESCGLE theory predicts sub-$T_g$ exothermic relaxation, superheated glass plateaus, endothermic devitrification overshoots, and Moynihan's reduced specific-heat superposition $\tilde{C}_p$ vs $T(t)/T'_f(q)$. 

\begin{acknowledgments}
This work was supported by the Secretar{\'i}a de Ciencia, Humanidades, Tecnolog{\'i}a e Innovaci{\'o}n (SECIHTI) through Postdoctoral Fellowship Grant No. 11200/320/2022, the Laboratorio Nacional de Ingenier{\'i}a de la Materia Fuera de Equilibrio (LANIMFE), and Monte Caldera Technologies.
\end{acknowledgments}

\appendix
\section{Self-Consistent Dynamic Closure and Asymptotic Non-Ergodicity Equations}\label{app:SCGLE_Closure}

To close the inner-clock trajectory $u(t) = \int_0^t b(t')\,\mathrm{d}t'$ self-consistently, at every instant $u$ (and corresponding instantaneous structure factor $S(k; u)$), the NESCGLE framework solves the coupled equations for the collective and self intermediate scattering functions $F(k, \tau_\alpha; u)$ and $F_S(k, \tau_\alpha; u)$ in the Laplace domain~\cite{Juarez2007, Pedro2010_2, Peredo2024}:
\begin{align}
F(k, z; u) &= \frac{S(k; u)}{z + \dfrac{k^2 D^0 S^{-1}(k; u)}{1 + \lambda(k)\,\Delta\zeta^*(z; u)}}, \label{eq:Fk_Laplace} \\
F_S(k, z; u) &= \frac{1}{z + \dfrac{k^2 D^0}{1 + \lambda(k)\,\Delta\zeta^*(z; u)}}, \label{eq:Fsk_Laplace}
\end{align}
coupled to the time-dependent friction memory kernel
\begin{equation}\label{eq:Delta_zeta}
\Delta\zeta^*(\tau_\alpha; u) = \frac{D^0}{36\pi\phi} \int_0^\infty \mathrm{d}k\,k^4 \left[\frac{S(k; u) - 1}{S(k; u)}\right]^2 F(k, \tau_\alpha; u)\,F_S(k, \tau_\alpha; u),
\end{equation}
where $\lambda(k) \equiv [1 + (k/k_c)^2]^{-1}$ is an interpolation function with cutoff wavevector $k_c = 1.305\,(2\pi/\sigma)$ calibrated to the hard-sphere dynamic arrest transition~\cite{Juarez2007, LuisEnrique2013}. The instantaneous collective mobility entering the inner clock~\eqref{eq:u_def} is then determined by
\begin{equation}\label{eq:b_from_zeta}
b(u) = \left[ 1 + \int_0^\infty \Delta\zeta^*(\tau; u)\,\mathrm{d}\tau \right]^{-1}.
\end{equation}

Whether an instantaneous structure factor $S(k; u)$ belongs to the ergodic fluid domain or the dynamically arrested glass domain is governed by the asymptotic ($\tau_\alpha \to \infty$) fixed-point equation for the squared cage localization length $\gamma(u) \equiv \lim_{\tau_\alpha\to\infty} \langle |\Delta\bm{r}(\tau_\alpha)|^2\rangle / 6$~\cite{Juarez2007, Pedro2010_2}:
\begin{equation}\label{eq:gamma_fixed_point}
\frac{1}{\gamma(u)} = \frac{1}{6\pi^2\overline{n}} \int_0^\infty \mathrm{d}k\,k^4\,\frac{\left[S(k; u) - 1\right]^2 \lambda^2(k)}{\left[\lambda(k)\,S(k; u) + k^2\gamma(u)\right]\left[\lambda(k) + k^2\gamma(u)\right]}.
\end{equation}
So long as the only solution of Eq.~\eqref{eq:gamma_fixed_point} is $\gamma(u) = \infty$ ($1/\gamma = 0$), the system remains ergodic with $b(u) > 0$. Once the structural trajectory $S(k; u)$ crosses the critical surface in state space where Eq.~\eqref{eq:gamma_fixed_point} bifurcates to a finite localization length $\gamma^a(q) < \infty$ ($1/\gamma^a > 0$), $b(u)$ vanishes as $u \to u^a(q)$, permanently arresting the non-equilibrium structure at $S^a(k; q) \equiv S(k; u = u^a(q))$.

Numerically, at each discrete inner-clock coordinate $u$ along a given thermal protocol, the self-consistent system of coupled integro-differential equations~\eqref{eq:Fk_Laplace}--\eqref{eq:Delta_zeta} for $F(k, \tau_\alpha; u)$, $F_S(k, \tau_\alpha; u)$, and $\Delta\zeta^*(\tau_\alpha; u)$ is solved in the correlation-time domain $\tau_\alpha$ using the open-source Julia package \texttt{ModeCouplingTheory.jl} developed by Pihlajamaa \textit{et al.}~\cite{Pihlajamaa2023}. The resulting memory kernel $\Delta\zeta^*(\tau_\alpha; u)$ is integrated via Eq.~\eqref{eq:b_from_zeta} to evaluate the instantaneous collective mobility $b(u)$, while the outer non-equilibrium evolution---comprising the piece-wise thermal protocol~\eqref{eq:step_u_def}--\eqref{eq:S_TNM_exact}, the adaptive mapping between laboratory waiting time $t$ and the inner clock $u(t)$, the fixed-point bifurcation solver for $\gamma(u)$ in Eq.~\eqref{eq:gamma_fixed_point}, and the thermodynamic and calorimetric post-processing---is implemented in our custom Julia framework.

\section{Geometric and Integral Constructions of the Calorimetric Glass Transition Temperature upon Cooling}\label{app:TgConstructions}

\begin{figure}[t!]
    \centering
    \includegraphics[width=0.45\textwidth]{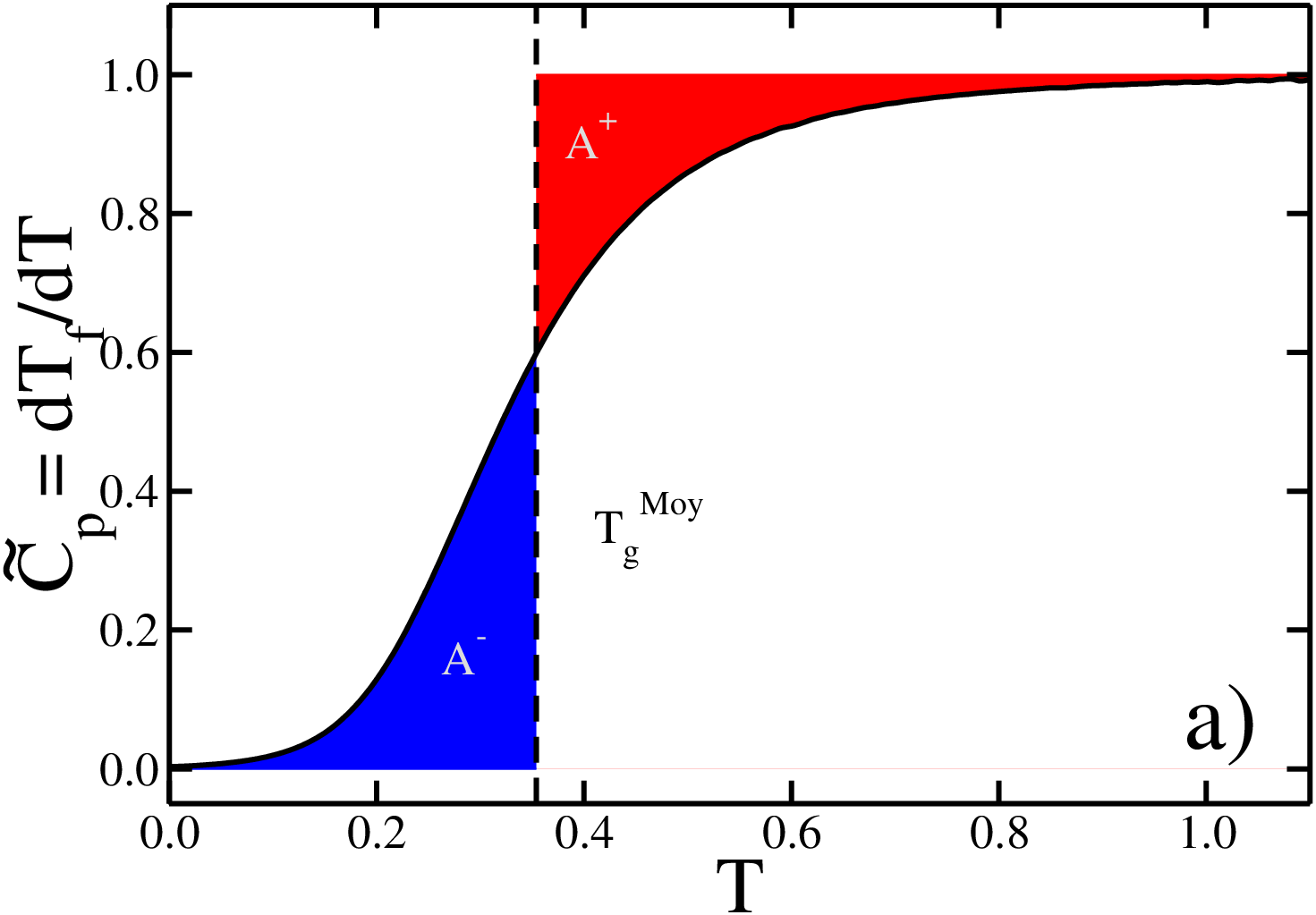}
    \includegraphics[width=0.45\textwidth]{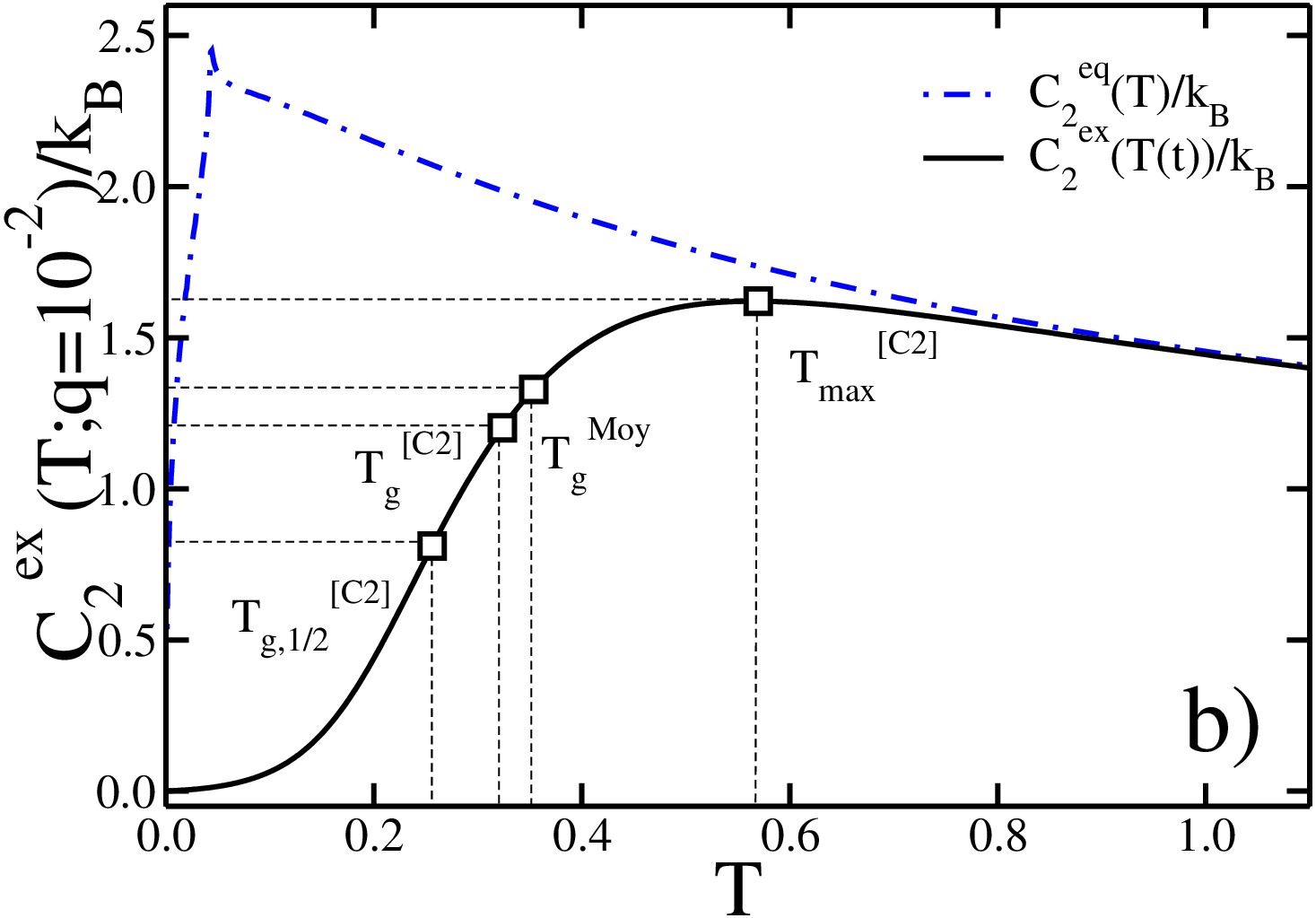}
    \caption{
    \textbf{(a)} Moynihan equal-area construction on $\tilde{C}_p^{\text{cool}}(T) = \mathrm{d}T_f/\mathrm{d}T$ [Eq.~\eqref{eq:Tg_calorimetric_integrals}]: the step at $T_g^{\text{Moy}} = T'_f = 0.354$ balances the deficit ($A^+$) and residual ($A^-$) areas [applying the same equal-area balance to $C_2^{\text{ex}}(T; q)/C_2^{\text{ex,eq}}(T)$ yields $T_g^{[C_2]} = 0.323$]. \textbf{(b)} Differential landmarks on $C_2^{\text{ex}}(T; q)/k_B$ versus $C_2^{\text{ex,eq}}(T)/k_B$: the onset peak $T_{\max}^{[C_2]} = 0.570$ ($C_{2,\max}^{\text{ex}}/k_B = 1.62$) and the half-step midpoint $T_{g,1/2}^{[C_2]} = 0.256$ ($C_2^{\text{ex}}/k_B = 0.81$).}
    \label{fig:app_Tg_constructions}
\end{figure}

Figure~\ref{fig:app_Tg_constructions} illustrates the integral and differential operational constructions of $T_g(q)$ synthesized in Sec.~\ref{sec:CalorimetricTg} (Fig.~\ref{fig:calorimetric_Tg}) for a representative cooling ramp at $q = 10^{-2}$ ($\phi = 0.70$, $T_c = 0.2287$, $M = 1000$).

\begin{enumerate}
    \item \textbf{Integral Equal-Area Constructions ($T_g^{\text{Moy}}$ and $T_g^{[C_2]}$) [Fig.~\ref{fig:app_Tg_constructions}(a)]:} Replacing the sigmoidal cooling curve $\tilde{C}_p^{\text{cool}}(T) = \mathrm{d}T_f/\mathrm{d}T$ between $T_i = 10.0$ ($\tilde{C}_p^{\text{cool}} = 1$) and $T_{\text{end}} = 10^{-3}$ ($\tilde{C}_p^{\text{cool}} = 0$) by an ideal step $\Theta(T - T_g^{\text{Moy}})$ that conserves $\Delta T_f$ equates the deficit area above $T_g^{\text{Moy}}$ ($A^+$) to the residual area below it ($A^-$):
    \begin{equation}\label{eq:app_Moynihan_cooling_balance}
    \mathcal{A}_{\text{deficit}}^{\text{cool}} \equiv A^+ \equiv \int_{T_g^{\text{Moy}}}^{T_i} \left[1 - \tilde{C}_p^{\text{cool}}(T; q)\right]\mathrm{d}T = \int_{T_{\text{end}}}^{T_g^{\text{Moy}}} \tilde{C}_p^{\text{cool}}(T; q)\,\mathrm{d}T \equiv \mathcal{A}_{\text{residual}}^{\text{cool}} \equiv A^-.
    \end{equation}
    Using $\tilde{C}_p^{\text{cool}} = \mathrm{d}T_f/\mathrm{d}T$ with $T_f(T_i; q) = T_i$ and $T_f(T_{\text{end}}; q) = T'_f(q) \gg T_{\text{end}}$ yields Eq.~\eqref{eq:Tg_calorimetric_integrals}, $T_g^{\text{Moy}}(q) = T'_f(q) = 0.354$, coinciding with the intersection of the glass plateau $T_f = T'_f(q)$ and the equilibrium line $T_f = T$ [Fig.~\ref{fig:fictive_temperature}(a)]. Applying the same equal-area balance ($A^+ = A^-$) to the two-body ratio $C_2^{\text{ex}}(T; q)/C_2^{\text{ex,eq}}(T)$ [Eq.~\eqref{eq:Tg_calorimetric_integrals}] gives $T_g^{[C_2]}(q = 10^{-2}) = 0.323$, which is slightly lower than $T_g^{\text{Moy}}(q)$ because $s_2(T; q)/k_B$ integrates over faster-relaxing modes $k \neq k_p$ [Fig.~\ref{fig:fictive_temperature}(b)].

    \item \textbf{Differential Calorimetric Landmarks ($T_{\max}^{[C_2]}$ and $T_{g,1/2}^{[C_2]}$) [Fig.~\ref{fig:app_Tg_constructions}(b)]:} On the unnormalized trace $C_2^{\text{ex}}(T; q)/k_B$, the departure from the rising equilibrium curve $C_2^{\text{ex,eq}}(T)/k_B$ produces a maximum $C_{2,\max}^{\text{ex}}/k_B = 1.62$ at the onset peak $T_{\max}^{[C_2]} = 0.570$. Below $T_{\max}^{[C_2]}$, the point where $C_2^{\text{ex}}/k_B$ drops to half its peak height ($0.81$) defines the half-step midpoint $T_{g,1/2}^{[C_2]} = 0.256$, which together with $T_{\max}^{[C_2]} = 0.570$ brackets $T_g^{\text{Moy}} = 0.354$ and $T_g^{[C_2]} = 0.323$.
\end{enumerate}

\bibliography{references}

\end{document}